\pdfoutput=1
\newif\ifreview
\reviewfalse
\ifreview
  \documentclass[review,12pt]{elsarticle}
\else
  \documentclass[final,5p,times,twocolumn]{elsarticle}
\fi
\usepackage{amssymb, amsthm, gensymb}
\usepackage[separate-uncertainty=true, parse-numbers=false]{siunitx}
\usepackage[colorlinks = {false}, pdfborder={0 0 0}]{hyperref}
\hypersetup{
  colorlinks=true,
  citecolor=blue,
  linkcolor=blue,
  urlcolor=blue
}
\usepackage{xurl} %% allow long URLs to break across lines (Data availability GitHub URL)
\usepackage{graphicx}
\usepackage{lineno} %% line numbers for review; activated only in review mode via \ifreview\linenumbers\fi
\usepackage{adjustbox}  %% scale wide tables to fit (no cut-off), kept upright
\usepackage{changepage} %% let wide tables use the full page width in single-column review
\usepackage{setspace}   %% \singlespacing for tables (the review body is double-spaced)
\ifreview
  \newenvironment{widetable}{\begin{table}[htbp]\centering}{\end{table}}
  \newcommand{\fitwide}[1]{\makebox[\textwidth]{\adjustbox{max totalheight=0.9\textheight,center}{\linespread{1}\small#1}}}
\else
  \newenvironment{widetable}{\begin{table*}[h!]}{\end{table*}}
  \newcommand{\fitwide}[1]{\adjustbox{max width=\textwidth,max totalheight=\textheight,center}{\linespread{1}\small#1}}
\fi
\usepackage{subfiles}
\usepackage{tabularx}
\usepackage{booktabs}
\usepackage{array}
\usepackage[T1]{fontenc} 
\usepackage{makecell}

\newcolumntype{L}[1]{>{\raggedright\arraybackslash}p{#1}}
\newcolumntype{C}[1]{>{\centering\arraybackslash}p{#1}}
\newcolumntype{R}[1]{>{\raggedleft\arraybackslash}p{#1}}

\usepackage{titlesec}
\usepackage{float}
\usepackage{wrapfig}
\usepackage{bm}
\usepackage{subcaption}
\usepackage{glossaries}
\usepackage{csquotes}
\usepackage{multirow}
\usepackage[justification=centering]{caption}
\usepackage{adjustbox}
\usepackage{comment}
\usepackage{xfrac}
\usepackage{fancyhdr}
\usepackage{multicol}
\usepackage{cleveref}
\usepackage[normalem]{ulem}
\usepackage{enumitem}
\usepackage{framed}

\usepackage{caption}
\usepackage{xcolor}

\makeatletter

\newrobustcmd*{\nobibliography}{%
 \@ifnextchar[%]
  {\blx@nobibliography}
  {\blx@nobibliography[]}}

\def\blx@nobibliography[#1]{}

\appto{\skip@preamble}{}

\makeatother

\journal{}

\makeatletter
\def\ps@pprintTitle{%
  \let\@oddhead\@empty
  \let\@evenhead\@empty
  \def\@oddfoot{\reset@font\hfil\thepage\hfil}
  \def\@evenfoot{\reset@font\hfil\thepage\hfil}
}
\makeatother

\fancypagestyle{pprintTitle}{%
  \fancyhf{}%
  \fancyfoot[C]{\thepage}%
  \fancyfoot[L]{\footnotesize\itshape Preprint.}%
}

\begin{document}

\begin{frontmatter}


%%       addressline={},
%%       city={},
%%       postcode={},
%%       state={},
%%       country={}}

\affiliation[aff_NM]{organization={University of Stuttgart, Institute of Space Systems},
      addressline={Pfaffenwaldring 29}, 
      city={Stuttgart},
      postcode={70569}, 
    %   state={},
      country={Germany}}

\affiliation[aff_WvL]{organization={Aliena Pte Ltd.},
      addressline={6 Ubi View}, 
      city={Singapore},
      postcode={408544}, 
     %  state={},
      country={Singapore}}

\affiliation[aff_AH]{%
 organization={University of Luxembourg},%
 addressline={Interdisciplinary Centre for Security, Reliability and Trust (SnT)},%
 city={Luxembourg},%
 postcode={L-1359},%
 % state={},%
 country={Luxembourg}%
}

\affiliation[I4IS]{organization={Initiative for Interstellar Studies},
      addressline={27/29 South Lambeth Road}, 
      city={London},
      postcode={SW8 1SZ}, 
     %  state={},
      country={UK}}

\title{High-temperature photovoltaics for solar-electric Oberth maneuvers: ton-class payload feasibility for interstellar-precursor missions}

\author[aff_NM,I4IS]{Nadim Maraqten\corref{cor1}}
\ead{maraqtenn [at] irs.uni-stuttgart.de}

\author[aff_WvL]{Willem van Lynden}
\author[I4IS]{Carlos Gómez de Olea Ballester}
\author[aff_AH,I4IS]{Andreas M. Hein}

\cortext[cor1]{Corresponding author.}
\begin{abstract}

In-situ exploration beyond the giant planets remains rare because timely Solar System escape demands very high specific orbital energy, which existing concepts typically achieve only with small payloads, super-heavy launchers, or nuclear-powered propulsion. Motivated by laboratory demonstrations of high-intensity, high-temperature (HIHT) solar cells operating near $400\,^{\circ}\mathrm{C}$, we assess a solar-electric Oberth maneuver that concentrates thrust near a $0.3\,\mathrm{AU}$ perihelion. Evolutionary steering optimisation indicates that an expendable Falcon Heavy could deliver ton-class payloads to $200\,\mathrm{AU}$ within 25 years if HIHT power systems reach specific powers about 10\% above present-day conventional levels with a Jupiter gravity assist, or about twice those levels on a direct trajectory, under the stated assumptions. The gain stems from a threefold increase in specific orbital energy for the same $\Delta v$ compared with a $1\,\mathrm{AU}$ spiral. These results suggest HIHT photovoltaics could shift from survival hardware to propulsion-enabling technology for high-energy deep-space missions.

\end{abstract}


\end{frontmatter}

\ifreview\linenumbers\fi


\section{Introduction}

Despite decades of near-Earth exploration, the outer heliosphere and heliopause remain sparsely sampled by in-situ missions, limiting our understanding of large-scale heliospheric structure, solar–interstellar coupling, and outer-heliosphere particle populations. Only a small number of spacecraft have achieved the high-energy Solar System escape trajectories required for timely in-situ measurements in these regions. This leaves key aspects of the heliosphere–interstellar-medium interaction at heliocentric distances on the order of 200~AU largely uncharacterised \cite{brandt2022interstellar,brandt2023future,linsky2023lies,erikssonhelio2050}. Addressing this gap is a major scientific and exploration goal which requires mission architectures that deliver high Solar System escape energy while retaining meaningful payload capability.

This limited sampling is primarily driven by the mismatch between the specific orbital energy required for timely Solar System escape and the practical limits of current propulsion systems. This energy gap restricts both achievable payload mass and overall mission duration. We therefore optimise allocatable science payload mass $m_{pl}$, interpreted here as the instrument suite and payload-specific hardware. Payload capability is treated as margin in sensitivity trades against EPS specific power and spacecraft structural-mass-fraction assumptions.

\begin{figure*}[t]
  \centering
  \includegraphics[width=\textwidth]{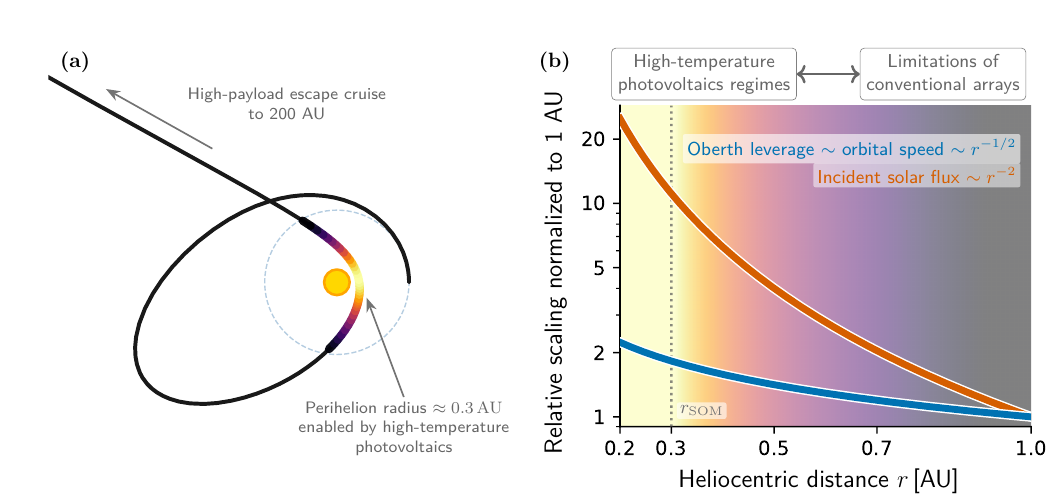}
  \caption{\textbf{Solar-electric propulsion solar Oberth maneuver concept and near-Sun scaling.}
  \textbf{(a)} Schematic of the SEP--SOM architecture, in which solar-electric propulsion delivers the spacecraft to a low perihelion and thrust is applied near perihelion to increase delivered specific orbital energy, enabling a high-payload escape cruise to $\sim$200~AU.
  \textbf{(b)} Relative scaling with heliocentric distance $r$, normalised to unity at 1~AU, of the perihelion orbital-speed leverage (the orbital speed that sets the Oberth energy gain per unit $\Delta v$, $\propto r^{-1/2}$) and the incident solar flux ($\propto r^{-2}$). The vertical marker indicates the reference perihelion radius $r_{\mathrm{SOM}}\approx0.3$~AU considered in this study.}
  \label{fig:sep_som_overview}
\end{figure*}

Conventional chemical propulsion, while reliable and well established, provides limited payload capacity and typically requires costly super-heavy-lift launchers. For example, published estimates place SLS Block~1B production cost at \$$2.5$~billion (excluding systems engineering and integration) \cite{NASA_OIG_2023_SLS}, whereas commercial heavy-lift procurement is reported at the order of \$$0.15$~billion for selected missions \cite{nasaGOESU2023}. Multiple stages or complex gravity-assist sequences are also commonly required to reach the high heliocentric velocities needed for deep-space exploration~\cite{hopkins2015propulsion}. The Voyager missions, for example, have provided key scientific insights beyond 130~AU \cite{Richardson2022OuterHeliosphere, Fuselier2020HeliopauseReconnection}, but their trajectories relied on the rare ``Grand Tour'' alignment of the outer planets \cite{Uri2017VoyagersGrandTour} that occurs only every $\sim$175~years \cite{butrica1998voyager}. Even with this favourable geometry, both spacecraft are now limited by asymptotic speeds of less than $4~\mathrm{AU\,year^{-1}}$ and aging systems, restricting further in-situ exploration. 

These limitations motivate architectures that increase delivered specific orbital energy without relying on new propulsion physics. Here we examine a Solar Oberth Maneuver (SOM) enabled by high-power electric propulsion and laboratory-demonstrated high-intensity, high-temperature (HIHT) photovoltaics. Executing $\Delta v$ near perihelion increases the gain in delivered specific orbital energy per unit $\Delta v$. Prior solar Oberth studies largely rely on impulsive chemical or solar-thermal stages with heavy thermal protection~\cite{hopkins2015propulsion,benkoski2023combined} or on high-performance solar sails~\cite{davoyan2021photonic}, leading either to high parasitic mass / launch demands or to kilogram-scale payloads. Existing solar Oberth concepts (see Table~\ref{tab:som_literature}) have not demonstrated ton-class payload delivery to $\sim$200 AU within 25 years under non-nuclear propulsion assumptions and commercial launch mass limits, motivating a solar electric variant.

Electric propulsion (EP) offers high $I_{\mathrm{sp}}$ and efficient propellant use~\cite{mazouffre2016electric}. In principle, the steep rise in solar irradiance toward perihelion can increase available power, making SEP--SOM attractive for outer Solar System and interstellar-precursor missions~\cite{genovese2023advanced,maraqten2024advanced}. However, prior solar electric Oberth concepts~\cite{bering2021solar} have been strongly constrained by thermal limits of conventional (non-HIHT) solar cells. To avoid cell degradation, these studies typically impose conservative minimum perihelia of $r_{\mathrm{SOM}} \approx 0.5$--0.7~AU \cite{bering2021solar, ohndorf2011flight}, where both perihelion velocity and irradiance are only modestly increased, yielding a correspondingly muted Oberth contribution and power level of only a few solar constants. These thermally constrained architectures therefore typically yield payloads of only tens to a few hundred kilograms to heliopause-like distances. This bottleneck is best seen by contrasting the near-Sun scalings of orbital-speed leverage and available solar power, alongside the SEP--SOM architecture (Fig.~\ref{fig:sep_som_overview}).

Recent advances in HIHT solar cell technology suggest a route to relaxing this bottleneck. Laboratory work by Perl \textit{et al.}\ \cite{perl2016measurements,Perl2018} and related studies \cite{sun2016thermal} have demonstrated that modern multi-junction cells can sustain efficient operation at temperatures approaching 400$^\circ$C under concentrated illumination in laboratory conditions. Building on these results, this study assumes HIHT survivability at $T \approx 400^\circ$C for the reference near-Sun operating condition and varies EPS specific power in the sensitivity analysis. For the reference trajectory optimisation, a system-level EPS specific power of $\alpha_{\mathrm{EPS}} = 200~\mathrm{W\,kg^{-1}}$ at 1~AU (including power processing units, PPU) is used as an illustrative high-performance anchor, motivated by published array-level development targets for deployable solar arrays (see Section~\ref{subsec:EPS_literature_review}). This baseline is used to quantify architectural leverage, not as a required feasibility threshold or near-term forecast; Section~\ref{subsec:sensitivity_analysis} then maps payload feasibility across lower and intermediate $\alpha_{\mathrm{EPS}}$ values.

Given these technology assumptions, the present work assesses a high-payload SEP--SOM architecture at $r_{\mathrm{SOM}} \approx 0.3$~AU, a near-Sun regime that array thermal limits had closed to solar-electric propulsion and that HIHT solar arrays could now open. We quantify the delivery of ton-class payloads to the heliopause ($\sim$200~AU) within 25~years on a commercial launcher (here: expendable Falcon Heavy), as an alternative to SLS-class options. Building on a preliminary architectural assessment~\cite{maraqten2024feis}, we document and validate the optimisation framework, derive payload-scaling relations and sensitivity maps, and benchmark the architecture against nuclear electric and impulsive Oberth concepts, contextualising related analytic perihelion estimates for advanced SEP probes~\cite{genovese2023advanced}.

To assess low-perihelion SEP--SOM as a high-payload solar-escape architecture, we make five main contributions: (i) an architecture-level SEP--SOM baseline at $r_\mathrm{SOM}\approx0.3~\mathrm{AU}$ that quantifies payload delivery under lumped subsystem assumptions and evaluates Falcon Heavy feasibility; (ii) a synthesis of prior solar Oberth concepts and near-Sun power constraints, distilling design drivers for low-perihelion SEP--SOM; (iii) the development, verification, and benchmark validation of the SOMBRERO (Solar Oberth Maneuver By Realisation of EvolutionaRy Optimisation) framework for planar SEP--SOM steering; (iv) the derivation and application of generalised payload-scaling relations and sensitivity maps for key parameters (EPS specific power and structural mass ratio); and (v) an interpretation of results positioning the HIHT-enabled SEP--SOM as a high-payload solar-escape architecture, contrasted with impulsive Oberth and nuclear electric approaches.

%=======================================================================
\section{Background and Literature Review}
\label{sec:background}
This section summarises prior SOM studies using diverse propulsion concepts. We then discuss recent advances in high-temperature solar cells, and the engineering and TRL gaps that must be closed to mature them into a flight-qualified, lightweight panel. 

\subsection{Solar Oberth Maneuver Theory}\label{subsec:solar_oberth_theory}

Let \(\varepsilon\) denote heliocentric specific orbital energy, \(\Delta\varepsilon\) its change due to propulsion, \(r\) the heliocentric distance, \(\Delta v\) a propulsive velocity increment, and \(I_{sp}\) the specific impulse. Let \(\mathbf{v}(t)\) be the heliocentric velocity and \(v(t)=\|\mathbf{v}(t)\|\) its magnitude; \(a\) denotes the heliocentric semi-major axis and \(r_p\) the perihelion radius. Maximising the Sun–escape speed is equivalent to maximising \(\varepsilon\).
Because the onboard \(\Delta v\) is limited, it is useful to maximise \(\Delta\varepsilon\) per unit \(\Delta v\). The Oberth effect states that a prograde burn yields a larger \(\Delta\varepsilon\) when executed at high \(v\) (deep in the potential well, i.e., small \(r\)); for the Sun this is termed a Solar Oberth Maneuver (SOM) \cite{oberth1929}.

In the impulsive limit, a prograde \(\Delta v\) changes specific orbital energy by \(\Delta \varepsilon = v\,\Delta v + \tfrac{1}{2}(\Delta v)^2\), highlighting why applying \(\Delta v\) near perihelion (high \(v\)) is advantageous. In practice, chemical stages can concentrate \(\Delta v\) near perihelion but are typically constrained in total \(\Delta v\) by \(I_{sp}\).

Electric propulsion systems, on the other hand, implement \(\Delta v\) over finite time, so an integral form needs to be considered:
\begin{equation}\label{eq:oberth_cont_general}
 \Delta\varepsilon
 \;=\;
 \int_{t_0}^{t_f} \mathbf{v}(t)\!\cdot\!\mathbf{a}_T(t)\,\mathrm{d}t
 \;=\;
 \int_{t_0}^{t_f} v(t)\,\frac{F_T(t)}{m_{\mathrm{sc}}(t)}\,\cos\!\theta_T(t)\,\mathrm{d}t,
\end{equation}

where $[t_0, t_f]$ is the thrust arc, $v(t)$ is the orbital speed, $F_T(t)$ is the thrust magnitude, $m_{\mathrm{sc}}(t)$ is the instantaneous spacecraft mass, and $\theta_T(t)$ is the angle between the thrust and velocity vectors. The integrand $v(t)F_T(t)/m_{\mathrm{sc}}(t)\,\cos\theta_T(t)=v(t)\,a_{\parallel}(t)$, where $a_{\parallel}(t)\equiv [F_T(t)/m_{\mathrm{sc}}(t)]\cos\theta_T(t)$ is the thrust acceleration component along the velocity, is the specific mechanical power delivered by propulsion; maximising its time integral maximises $\Delta\varepsilon$.

For a SEP--SOM, a lower perihelion is attractive because it increases (i) the local orbital speed and thus the energy leverage per unit \(\Delta v\), and (ii) the available solar electric power (and hence attainable thrust acceleration) at small \(r\), allowing more of the thrust work to be delivered during the high-\(v\) perihelion arc. For illustration, for \(a=3\)~AU, reducing \(r_p\) from 0.7 to 0.3~AU increases perihelion speed from \(\approx 47.3\) to \(75.0~\mathrm{km\,s^{-1}}\) (+58\%). Together, lower perihelion increases both the speed \(v(t)\) and the attainable thrust acceleration component along the velocity, \(a_{\parallel}(t)\), thereby increasing the integrand in \eqref{eq:oberth_cont_general} and motivating thrust work concentrated near perihelion.

\subsection{Solar Oberth Maneuver Literature Review}\label{subsec:solar_oberth_literature_review}

%%%%%
Prior Solar Oberth Maneuver studies span impulsive staging, sail-based concepts, and solar electric architectures. Across these proposals, high escape energy is typically traded off against parasitic mass, conservative perihelia, or both, which limits delivered science capability in the reported reference cases. Table~\ref{tab:som_literature} summarises representative studies and reproduces payload figures as stated in the cited sources, with the payload definition indicated in the table footnotes. The proposals fall into three categories:

\begin{enumerate}
  \item \textbf{High-Thrust Impulsive SOMs:} This category, which includes chemical propulsion, Solar Thermal Propulsion (STP), and Nuclear Thermal Propulsion (NTP), seeks to maximise the Oberth effect by adopting extreme-proximity perihelia, typically ranging from 2--11 solar radii (\,$R_\odot$) \cite{hopkins2015propulsion, benkoski2023combined, lyman2001_stp_interstellar}. The resulting thermal environment necessitates massive, robust thermal-protection systems (TPS). The parasitic mass of this heat shield and/or active cooling hardware curtails the available \emph{science payload} or residual payload margin. Reported science payloads fall in the tens-of-kilograms range in some studies \cite{sauder2021system}, and others note tight residual-mass margins once staging and any required perihelion heat shielding are accounted for \cite{hibberd2021project}. These architectures also often depend on high-cost, heavy-lift launchers such as the SLS to provide the requisite launch energy \cite{hopkins2015propulsion, benkoski2023combined, hibberd2022can}. At the trajectory level, such low-perihelion, high-thrust SOMs are highly sensitive to injection and navigation errors. Once the impulsive burn has been executed, any mismatch between the targeted and achieved perihelion state maps into large errors in the outgoing hyperbola, leaving little scope for post-SOM correction. When these impulsive SOMs are embedded in additional gravity-assist sequences (e.g. JGA), the overall mission-design and operations complexity, as well as sensitivity to launch-window phasing, increases.

  \item \textbf{High-Performance Solar Sails:} While capable of very high escape velocities ($20-60~\mathrm{AU\,year^{-1}}$) using solar radiation pressure at close perihelia (5--11\,$R_\odot$), solar sails are mass-limited. Their performance is dictated by an extreme area-to-mass ratio and very large deployable structures, which typically restricts the delivered dry spacecraft mass to the kilogram-to-tens-of-kilograms class \cite{davoyan2021photonic, bailer2021sun, davoyan2024extreme}.
  
  \item \textbf{Thermally constrained solar-electric SOMs (high-perihelion):} This category, the most relevant to the present study, has historically been power-system-constrained \cite{Kluever1997_SEP_Interstellar}. To protect conventional solar arrays from thermal damage, prior solar-electric SOM studies, including work explicitly framed as a ``solar electric Oberth maneuver'' \cite{bering2021solar}, are confined to conservative perihelia of $r_{\mathrm{SOM}} \approx 0.5$--0.7\,AU \cite{bering2021solar, ohndorf2011flight,Kluever1997_SEP_Interstellar}. This high-perihelion approach has two fundamental disadvantages: it blunts the Oberth leverage by limiting perihelion velocity, and it restricts the available solar intensity for the electric thrusters to only $\lesssim$4 solar constants. These thermally constrained concepts yield comparatively low \emph{science payloads} where reported (e.g., 35\,kg instrument mass) and do not provide a viable pathway to high-payload missions \cite{ohndorf2011flight,Kluever1997_SEP_Interstellar,loeb2011interstellar}.

\end{enumerate}

The literature does not yet establish a SOM architecture that supports a large allocatable mission payload. Impulsive concepts are limited by parasitic thermal-protection mass, while prior solar electric concepts are limited by thermally constrained, high-perihelion trajectories. Table~\ref{tab:som_literature} provides a quantitative summary of key assumptions, performance parameters, and limiting factors. For completeness, non-electric SOM implementations (chemical, solar sail, solar thermal, and nuclear thermal) are reviewed in \ref{app:non_ep_som_propulsion}. The remainder of this subsection focuses on electric-propulsion SOM architectures (Section~\ref{sec:electric_propulsion_SOM}), which form the closest antecedent to the present feasibility model.

\begin{widetable}
\caption{\textbf{Solar Oberth Maneuver mission metrics reported in the literature for benchmarking.} No surveyed concept demonstrates ton-class payload delivery to $\sim$200~AU within 25 years using non-nuclear propulsion and a commercial launch.}
\label{tab:som_literature}
\centering
\setlength{\abovecaptionskip}{4pt}\setlength{\belowcaptionskip}{2pt}
\ifreview\scriptsize\else\small\fi
\setlength{\tabcolsep}{3.5pt}
\renewcommand{\arraystretch}{1.0}
\singlespacing % review body is double-spaced; keep the table single-spaced (no-op in 2-column)

% Review: render at the natural column width (\scriptsize) and centre with \makebox so the
% table bleeds ~0.5 in into the margins -- the single-column text block is too narrow for this
% wide table at a readable size, and scaling it down further is too small. Two-column: scale
% proportionally to the full column-pair width. Column widths are hand-tuned so NO cell
% overflows: in particular the m_pl column holds "(limited)\textsuperscript{U}" /
% "(constr.)\textsuperscript{U}" and the v_inf column holds the "[AU year^-1]" unit header.
\makebox[\textwidth]{%
\begin{tabular}{@{}%
 L{1.0cm} % Prop. type ("Chem." etc.)
 L{2.8cm} % Reference
 L{2.0cm} % Launcher
 C{1.1cm} % C3 [km^2 s^-2]
 C{1.3cm} % Perihelion rp
 C{1.0cm} % m0 [kg]
 C{1.45cm} % mpl [kg] -- wide enough for "(limited)^U", "(constr.)^U"
 C{1.6cm} % v_inf [AU year^-1] -- wide enough for the unit header
 C{1.1cm} % t200AU [yr]
 L{3.5cm} % Performance / Key limitation
@{}}
\toprule

{\bfseries Prop. Type} &
{\bfseries Reference} &
{\bfseries Launcher Type} &
{\bfseries $\mathbf{C_3}$ [km$^{2}$\,s$^{-2}$]} &
{\bfseries $\mathbf{r_{SOM}}$ [AU]} &
{\bfseries $\mathbf{\ m_0\ }$ [kg]} &
{\bfseries $\mathbf{m_{pl}}$ [kg]$^{\dagger}$} &
{\bfseries $\mathbf{v_{\infty,\odot }}$ [$\mathbf{AU\,year^{-1}}$]} &
{\bfseries $\mathbf{t_{\text{200AU}}}$ [yr]} &
{\bfseries Performance / Key Limitation} \\

\midrule
Chem. &Hopkins et al.\ \cite{hopkins2015propulsion} & SLS B1 + JGA &
$\approx135$ & $\approx 0.05$\textsuperscript{c} & (N/A) & $\sim 380$\textsuperscript{D}\textsuperscript{a} &
$\approx 10$ & $\approx 20$ & \makecell[l]{100+ AU in 10 yrs\\Chem. stages only for SOM;\\otherwise SEP / Solar Sail
}. \\
Chem. &Hibberd et al.\ \cite{hibberd2022can} & SLS B2 + JGA &
$\approx100$ & (low) & (N/A) & 100\textsuperscript{D} &
$\approx 8.5$ & $\approx 24$ & $\sim$47 yrs to 400 AU. \\
Chem. &Hibberd \& Eubanks\ \cite{hibberd2026catching3iatlasusingsolar} & Starship B3 + JGA &
$\approx130$ & $\approx0.015$ & 16{,}450 -17{,}754 & 342-546\textsuperscript{D} &
$\approx 10-20$ & $\approx 13$ & \makecell[l]{General feasibility study.\\Heat shield mass\\not accounted for.} \\
\midrule
\makecell[l]{Solar \\Sail} & Davoyan et al.\ \cite{davoyan2021photonic} & (N/A) &
(N/A) & 0.05 & (N/A) & 10--50\textsuperscript{D} & $>20$ & $<10$ &
Requires $>10{,}000$\,m$^2$ sail. \\
\makecell[l]{Solar \\Sail} &
\makecell[l]{Liewer et al.\ \cite{liewer2000_nasa_interstellar_probe}} &
Delta II & (N/A) & $\approx 0.25$ & $\approx 246$ & 25\textsuperscript{I}& $\approx 15$ & $\approx 15$ & \makecell[l]{400\,m sail ($\sim 1~\mathrm{g\,m^{-2}}$).} \\
\makecell[l]{Solar \\Sail} & Bailer-Jones \cite{bailer2021sun} & (N/A) &
(N/A) & 0.05--0.53 & (N/A) & (N/A)\textsuperscript{U} & (N/A) & (N/A) &\makecell[l]{Theoretical trajectory study; \\ analyses maneuver efficiency.} \\
\makecell[l]{Solar \\Sail} & Davoyan et al.\ \cite{davoyan2024extreme} & (N/A) &
(N/A) & $<0.023$\textsuperscript{c} & 10--20 & 12--16\textsuperscript{D} & $\approx60$ & $\approx3$ & \makecell[l]{$\sim$1{,}000 AU in $\approx$17 yrs \\(concept).} \\
\makecell[l]{Laser \\Sail} & Hibberd et al.\ \cite{hibberd2020projectlyracatching1ioumuamua} & (N/A) &
(N/A) & $\approx 0.015$\textsuperscript{c} & $<\!1$ & $<\!1$\textsuperscript{D} & $\approx63.2$\textsuperscript{b} & $\approx3$ &
Gram–kg class (0.001$c$). \\
\midrule
STP & Shoji et al.\ \cite{shoji1992_solarthermal} & (N/A) &
(N/A) & $\approx 0.02$\textsuperscript{c} & (N/A) & (N/A)\textsuperscript{U}
 & $\approx 11.0$ & $\approx18$ &
Early conceptual study. \\
STP & Lyman et al.\ \cite{lyman2001_stp_interstellar} & JGA &
(N/A) & $\approx 0.016$\textsuperscript{c} & (N/A) & $\sim 50$\textsuperscript{D} & 20.0 & 10 &
750~kg SOM vehicle; limited by tank mass. \\
STP & Sauder et al.\ \cite{sauder2021system} & SLS B2 + GA\textsuperscript{d} &
(N/A) & $\approx0.015$\textsuperscript{c} & 478 & 36\textsuperscript{I} & 8.0–10.0 & $\approx20$--$25$ &
Various gravity assists. \\
STP & Benkoski et al.\ \cite{benkoski2023combined} & SLS + JGA &
165 & $\approx0.01$\textsuperscript{c} & 5{,}000 & 661.6\textsuperscript{D} & 15.0 & $\approx13$ &
\makecell[l]{Limited payload.} \\
\midrule
NTP & Hibberd \& Hein \cite{hibberd2021project} & SLS B2 &
(N/A) & (low) & \makecell[l]{1{,}300 -\\7{,}800} & (limited)\textsuperscript{U} & (N/A) & (N/A) &
\makecell[l]{Residual mass margin \\highly limited.} \\
NTP & Edwards et al.\ \cite{irvine2020design} & Various &
(N/A) & $\approx 0.05$\textsuperscript{c} & $\sim500$\textsuperscript{e} & (constr.)\textsuperscript{U} & $13.3-14.8$ & $20-33$ & Multiple launch vehicles and propellant options considered. \\
NTP & Scott D.\ \cite{scott2018analysis} & SLS B1 &
(N/A) & $\approx 0.05$\textsuperscript{c} & (N/A) & (constr.)\textsuperscript{U} & $\approx 9$ & $\approx22$ & Targets 500~AU. \\
\midrule
{EP} & {Kluever \cite{Kluever1997_SEP_Interstellar}} & {Titan IV + JGA} &
17.92 & {0.7} & {(N/A)} & {200\textsuperscript{D}} & $\approx7.6$ & 26.4 &
Thermally limited. \\
{EP} & Loeb et al.\ \cite{loeb2011interstellar}; Ohndorf et al.\ \cite{ohndorf2011flight} & Ariane 5 + JGA &
{45.1} & {0.7} & {1{,}692} & {35\textsuperscript{I}} & 8.0 & 25 & Thermally limited. REP\textsuperscript{f} stage required. \\
{EP} & \makecell[l]{Bering et al.\ \cite{bering2021solar}} & {\makecell[l]{Heavy Lift +\\SGA + NGA\textsuperscript{g}}} &
{25} & {0.5--0.66} & {$30{,}000$} & {(N/A)\textsuperscript{U}} & 8.0 & 25 &
\makecell[l]{1,000 AU in 125 yrs; \\thermally limited.} \\
\bottomrule
%% Footnotes as ONE compact, flowing row spanning the table (p{} ~= the column sum), so they
%% stay attached, share the table width, and travel/scale with the table -- they can never
%% detach or split, and the flowing form keeps the block on one page.
\multicolumn{10}{@{}p{16.8cm}}{\footnotesize\raggedright
$^{\dagger}$Superscripts indicate the payload definition.\quad
\textsuperscript{I}Science payload (instrument mass).\quad
\textsuperscript{D}Delivered dry spacecraft mass, excluding propellant and any jettisoned transfer/propulsion stage.\quad
\textsuperscript{U}Payload definition not specified or ambiguous in the cited source (or no mission payload value reported).\quad
\textsuperscript{a}Baseline bus $\sim$380\,kg; additional $\sim$300\,kg heat shield required.\quad
\textsuperscript{b}From 0.001$c$; $1c \approx 63{,}241~\mathrm{AU\,year^{-1}}$.\quad
\textsuperscript{c}Converted from value in \,$R_\odot$; \,$R_\odot \approx 0{.}00465$ AU.\quad
\textsuperscript{d}EGA (Earth Gravity Assist) + VGA (Venus Gravity Assist) + JGA (Jupiter Gravity Assist).\quad
\textsuperscript{e}"New Horizon's type" spacecraft.\quad
\textsuperscript{f}Radioisotope Electric Propulsion.\quad
\textsuperscript{g}SGA (Saturn Gravity Assist) + NGA (Neptune Gravity Assist).}\\
\end{tabular}}
\end{widetable}

\subsubsection{Electric Propulsion SOM}\label{sec:electric_propulsion_SOM}

Literature on electric propulsion solar Oberth maneuvers is limited, mainly due to thermal constraints on conventional solar arrays. Identified studies have been summarised in Table~\ref{tab:som_literature}.

Kluever~\cite{Kluever1997_SEP_Interstellar} analysed SEP and nuclear-electric interstellar-precursor missions to 200 AU across several launch vehicles, with the perihelion burn near 0.7~AU to maximise solar-array performance. For the Titan IV/Centaur SEP case with a Jupiter gravity assist, Kluever reports a 26.4-year transfer for a fixed 200~kg net spacecraft mass (payload plus basic structure). Loeb~\cite{loeb2011interstellar} and Ohndorf~\cite{ohndorf2011flight} proposed a smaller Ariane 5-launched mission (1,692~kg) requiring a JGA and Radioisotope Electric Propulsion (REP) stage to reach 200 AU in 25~years, leaving only 35~kg for scientific payload. Their proposal serves as reference for the paper at hand, as the developed neurocontrol algorithm is verified using their results. Finally, Bering et al.~\cite{bering2021solar} investigate solar and hybrid EP mission concepts to the Kuiper Belt and beyond. One trajectory uses Saturn and Neptune gravity assists, with the Neptune flyby setting up an Eris flyby and a subsequent coast to 1{,}000~AU in 125~years, an average of $8~\mathrm{AU\,year^{-1}}$. The concept places perihelion near 0.5~AU, and one illustrated case shows a closest solar approach of 0.657~AU. Their ice-giant modelling builds on a stated propulsion basis model that assumes argon operation ($I_{\mathrm{sp}}\approx 5{,}200$~s) until the spacecraft reaches 1~AU on the outbound leg, after which it switches to hydrogen ($I_{\mathrm{sp}}=32{,}800$~s). A refined variant includes jettison of depleted argon tankage near 1~AU and assumes a spacecraft launch mass of roughly 30~t and an Earth-relative hyperbolic excess speed of $v_{\infty,\text{Earth}}\approx 5~\mathrm{km\,s^{-1}}$, implying $C_3\approx 25~\mathrm{km^2\,s^{-2}}$. For the 1{,}000~AU case, delivered payload mass and EPS or array specific-power assumptions are not reported alongside the time-of-flight result. 

Literature that explicitly frames a low-perihelion, solar-powered EP thrust arc as an Oberth-type maneuver for Solar System escape remains limited. Conceptually adjacent strategies appear in the broader SEP mission literature for high-energy interplanetary transfers. These related concepts generally remain in a more conservative near-Sun regime and rely on explicit array thermal control. Zola~\cite{zola1969sep} presents early SEP probe concepts and shows a representative array power profile that is flattened below about $0.65~\mathrm{AU}$ by tipping the panel to maintain roughly constant equilibrium temperature. Rodgers and Brophy~\cite{rodgers2001tempo} propose the TEMPO concept for a fast Pluto flyby and note that the array is feathered for thermal control below $1~\mathrm{AU}$, with temperature margin down to about $0.6~\mathrm{AU}$. Ilin et al.~\cite{ilin2010vasimr} survey VASIMR mission strategies and include a reusable catapult concept that uses an inward arc to raise solar power and then accelerates near the Sun. Their example reaches $R_{\min}=0.452~\mathrm{AU}$ and assumes planar arrays to avoid overheating photovoltaic cells.
These studies are not treated as SEP--SOM escape analyses and are therefore not included in Table~\ref{tab:som_literature}. They support the broader idea that near-Sun solar electric propulsion can front-load energy for fast outer-planet transfers and other high-energy missions. They also help delineate the remaining gap addressed here, namely that pushing solar-powered EP thrusting to substantially smaller perihelia requires higher-temperature EPS and array capability.

\subsection{Electric Power System Literature Review}\label{subsec:EPS_literature_review}

The minimum perihelion for a SOM is primarily limited by thermal constraints. Close to the Sun, high operating temperatures due to the high-intensity solar irradiation can severely degrade solar cell performance \cite{dupre2017thermal,genovese2023advanced}, while the spacecraft structure itself must also withstand extreme heating. Thermal shielding can protect the spacecraft, but the solar panels must remain exposed to collect power for the propulsion system, making HIHT solar cells a critical enabling technology.

Only a few missions have flown close to the Sun. Helios 1 and 2 (1974–1976) reached 0.29 AU using silicon cells on angled panels with quartz mirrors covering half the array area to limit temperatures to $\sim$160$^{\circ}$C \cite{porsche1981helios}. Parker Solar Probe (2018) reached 0.044 AU using advanced multi-junction cells designed for $\sim$160$^{\circ}$C. Active cooling behind the heat shield kept the cells near this limit while the Sun-facing surfaces reached temperatures over $\sim$900$^{\circ}$C \cite{boca2013UV,raouafi2023parker}. The Solar Orbiter (2020) reaches 0.28 AU with triple-junction GaAs cells capable of $\sim$230$^{\circ}$C, supplemented by reflective coatings and panel tilting to limit solar absorption \cite{lyngvi2005solar,oberhuttinger2017simulating,marirrodriga2021solar}. Mercury orbiters such as Messenger and BepiColombo, as well as earlier Venus and Mercury missions like Mariner 10 and the Venera series, also faced near-Sun conditions requiring similar strategies.

These missions employ four main approaches to ensure solar panel survival: (i) using specialised high-temperature cells, (ii) reducing incident solar irradiation through tilting or shading, (iii) lowering panel absorption via reflective coatings or mirrors, and/or (iv) including active cooling. Active cooling adds mass, reducing payload, while shading or reflective coatings limit the power collected. To maximise power during a SOM and exploit the Oberth effect, it is therefore critical to capture as much solar flux as possible, highlighting HIHT solar cells as a central enabling technology. Notably, the present lumped-parameter mass budget omits a separate Parker Solar Probe–style dedicated heat shield and active-cooling hardware for the spacecraft bus. Instead, the bus is assumed to remain largely in the geometric shadow of the Sun-facing HIHT arrays during the perihelion arc; any residual shielding or thermal-control mass is subsumed into the structural mass fraction $\mu_s$ and not modelled explicitly.

\subsubsection{Advancements in High-Temperature Solar Cells}\label{subsub:HIHT_literature_review}

\begin{widetable}
\caption{\textbf{Summary of High-Intensity High-Temperature (HIHT) solar cell developments.} Reported temperatures refer to cell-level results. Array-level integration and system-level power-processing impacts are treated separately.}
\label{tab:hiht_cells}
\centering
\small
\setlength{\tabcolsep}{4pt}
\renewcommand{\arraystretch}{1.15}

% Column layout: adjust widths so total ≲ \textwidth
\fitwide{%
\begin{tabular}{@{}%
 L{3.5cm} % Material / Cell Type
 C{3.5cm} % Temperature
 L{6.5cm} % Application
 C{2.0cm} % Reference
@{}}
\toprule
\textbf{Material / Cell Type} &
\textbf{Sustained / Tested Temperature} &
\textbf{Intended Application} &
\textbf{Reference(s)} \\
\midrule

GaInP &
$\sim 400^\circ$C (predicted) &
Early HIHT concept for Mercury-, Venus-, and near-Sun exploration. &
\cite{scheiman1999high} \\

GaAs/AlGaAs concentrator cells &
$>500^\circ$C (survived to $800^\circ$C) &
High-intensity concentrator photovoltaics subjected to extreme thermal cycling. &
\cite{Spitzer1988} \\

SiC (6H--SiC) solar cells &
Up to $600^\circ$C &
Operation within $\sim$3 solar radii ($\approx 0.014$ AU); very high-temperature near-Sun missions. &
\cite{bailey2001silicon} \\

InGaP/InGaAs/Ge triple-junction &
Up to $\sim240^\circ$C &
Cells designed for near-Sun operation and high-flux illumination (up to $\sim$200$\times$ concentration). &
\cite{nishioka2005evaluation} \\

InGaP/GaAs/Ge triple-junction &
Up to $\sim230^\circ$C &
Near-Sun missions. &
\cite{brandt2013influence} \\

AlGaInP, GaAs, GaInP/GaAs dual-junction &
Up to $400^\circ$C &
HIHT laboratory cells approaching near-Sun operational requirements (up to $\sim$1{,}000$\times$ solar flux). &
\cite{perl2016measurements,Perl2018} \\

GaInP/GaAs (Venus-lander design) &
Up to $465^\circ$C &
Photovoltaic systems tailored for Venus’s extreme surface temperature environment. &
\cite{grandidier2020photovoltaic} \\

InGaN-based HIHT cells &
Up to $450^\circ$C &
Next-generation HIHT photovoltaics for extreme-temperature space environments. &
\cite{zhao2022toward} \\

\bottomrule
\end{tabular}}
\end{widetable}

HIHT solar cells have been studied for decades, motivated by missions to Mercury, Venus, and near-Sun exploration. The BepiColombo solar arrays are a recent flight-demonstrated case, using GaAs triple-junction cells (AZUR 3G28). These cells are validated only to 215$^{\circ}$C \cite{2011BCcells}. Similarly, Parker Solar Probe uses triple-junction cells, but relies on active cooling to maintain operating temperatures near 160$^{\circ}$C~\cite{boca2013UV}. While these missions represent the current state of flight-ready technology, research shows that laboratory and near-Sun designs can operate at far higher temperatures without such extensive thermal control. Table~\ref{tab:hiht_cells} presents a compilation of recent developments, showing the progression of HIHT photovoltaics toward practical near-Sun applications at temperatures exceeding \SI{400}{\celsius}.

These results are largely cell-level. Further development into full, flight-like panels may still require appropriate component shielding \cite{lindner2019solar}.

\subsubsection{Specific power}\label{subsub:specific_power_literature_review}
Beyond surviving the near-Sun environment, an equally important metric for solar-driven electric propulsion missions is the specific power $\alpha$, the power delivered per unit mass.
For the majority of commercially developed triple-junction cells, $\alpha_{\mathrm{cell}} < 800~\mathrm{W\,kg^{-1}}$, with efficiencies in the range of 20-30\% \cite{fatemi2000solar,MejiaEscobarAlgora2025_Joule_FlexiblePVAs}.

Significant progress has been made in further increasing cell-level $\alpha$. Commercial Off-The-Shelf (COTS) triple-junction cells have been commercially developed with specific powers up to $3{,}200~\mathrm{W\,kg^{-1}}$~\cite{MejiaEscobarAlgora2025_Joule_FlexiblePVAs} while many other advanced multi-junction and perovskite cells have achieved specific powers above $1{,}000~\mathrm{W\,kg^{-1}}$ ~\cite{nassiri2021high,ho2022deployment,tu2021perovskite}, with some exceeding $10{,}000~\mathrm{W\,kg^{-1}}$  \cite{kaltenbrunner2015flexible,kang2019perovskite}. 

Array assembly components beyond the solar cells (cover glass, structure, insulation, harness) reduce the array-level specific power $\alpha_{\text{array}}$ relative to $\alpha_{\mathrm{cell}}$; thermal-mitigation strategies (tilting, reflective optics, active cooling) can further reduce $\alpha_{\text{array}}$. If the solar array provides $\alpha_{\text{array}}$ and the power-processing unit (PPU) provides $\alpha_{\text{PPU}}$, the combined EPS performance is given by
\begin{equation}\label{eq:alpha}
  \alpha_{\text{EPS}} = \big(1/\alpha_{\text{array}} + 1/\alpha_{\text{PPU}}\big)^{-1}.
\end{equation}
Recent surveys of deployable arrays report array-level specific powers exceeding $100~\mathrm{W\,kg^{-1}}$ (e.g., iROSA) and $\sim 112~\mathrm{W\,kg^{-1}}$ for the Phoenix UltraFlex wing (flight heritage). Their successors are projected higher still: the roll-out (ROSA) architecture to $\sim 218~\mathrm{W\,kg^{-1}}$, and the scaled-up MegaFlex wing to $\sim 250~\mathrm{W\,kg^{-1}}$~\cite{MejiaEscobarAlgora2025_Joule_FlexiblePVAs}. Assuming comparable scaling, Perovskite panels might increase roughly tenfold, to $\gtrsim 2{,}000~\mathrm{W\,kg^{-1}}$. While further research is required to validate these projections and to develop heat-resistant panel structures, these advances suggest that combining HIHT capability with high-$\alpha$ cells may enable significantly more powerful electric propulsion systems operating much closer to the Sun than is currently feasible.

Present-day PPUs report $\alpha_{\text{PPU}}= 266~\mathrm{W\,kg^{-1}}$~\cite{jackson201713kw} (Lunar Gateway). Together with $\alpha_{\text{array}}=112~\mathrm{W\,kg^{-1}}$, Eq.~\eqref{eq:alpha} yields $\alpha_{\text{EPS}}\approx 78.8~\mathrm{W\,kg^{-1}}$ as an order of magnitude value for present-day solar electric EPS-level specific powers.

Beyond heritage hardware, multiple agency studies cite a development objective of $200~\mathrm{W\,kg^{-1}}$ for deployable solar-array specific power $\alpha_{\text{array}}$. The ESA High-Efficiency Solar Arrays study targets $200~\mathrm{W\,kg^{-1}}$ for concepts above $150$~kW \cite{ESA_HESA_Nebula}. A NASA programme description for MegaFlex cites a wing-level goal of $200~\mathrm{W\,kg^{-1}}$ at the beginning of mission near one astronomical unit \cite{NASA_TechPort_MegaFlex_9879}. Propulsion reviews discuss similar spacecraft-level projections over five to ten years \cite{Jovel2022}. At system level, Eq.~\eqref{eq:alpha} places $\alpha_{\text{EPS}}$ strictly below $\alpha_{\text{array}}$; closing the gap requires the PPU to outpace the array. Using the present-day example above, $\alpha_{\text{EPS}}=200~\mathrm{W\,kg^{-1}}$ under uniform scaling implies $\alpha_{\text{array}}\approx 284~\mathrm{W\,kg^{-1}}$ and $\alpha_{\text{PPU}}\approx 675~\mathrm{W\,kg^{-1}}$. The implied PPU value should be read as a stretch target, not a heritage-level value. For context, NASA GRC projected an optimised 14~kW-class Hall-thruster PPU at \(\gtrsim 330~\mathrm{W\,kg^{-1}}\) \cite{Pinero2015_HighPowerHallPPU}, while SiC-based developments target \(>2500~\mathrm{W\,kg^{-1}}\) at converter-stage level \cite{Reese2015_SiC_PPU_HiVHAC}.

\subsection{High Power Electric Propulsion Literature Review}\label{subsec:HPEP}

For the reference mission, an electric thruster with $I_{sp}=6{,}000$~s and $\eta=0.75$ is assumed. Gridded ion thrusters (GIT) have demonstrated these values in the laboratory \cite{polk2012high, leiter2014results}, while applied-field magnetoplasmadynamic thrusters (AF-MPD) approach them, with a 150~kW superconducting prototype reaching $I_{sp}\approx5{,}700$~s at $\eta\approx0.77$ \cite{zheng2021integrated}. Thrusters identified to meet or nearly meet the reference mission required values are listed in Table~\ref{tab:thrusters_big}.

\begin{widetable}
\caption{\textbf{Summary of high-specific-impulse electric propulsion thruster concepts and representative thruster models relevant for Solar Oberth Maneuver missions.} GIT denotes a gridded ion thruster. AF-MPD denotes an applied-field magnetoplasmadynamic thruster.}
\label{tab:thrusters_big}
\centering
\linespread{1}\small  % single spacing (review body is double-spaced; keeps the AF-MPD multirow bullets inside the rules)
\setlength{\tabcolsep}{4pt}
\renewcommand{\arraystretch}{1.05}

% Column layout consistent with the HIHT cell table
\fitwide{%
\begin{tabular}{@{}%
 L{1.5cm}  % Thruster Type ("AF-MPD" fits without overflow)
 L{3.7cm}  % Thruster Characteristics
 L{1.7cm}  % Model
 L{3.7cm}  % Model Performance / Characteristics
 C{1.8cm}  % Isp
 C{1.3cm}  % Efficiency
 C{1.1cm}  % Power
 L{1.5cm}  % References
@{}}
\toprule
\textbf{Type} &
\textbf{General Characteristics} &
\textbf{Model} &
\textbf{Performance Notes} &
\textbf{$I_{sp}$ [s]} &
\textbf{$\eta$ [-]} &
\textbf{Power [kW]} &
\textbf{References} \\
\midrule

% ----------------------- GIT -----------------------
\multirow{4}{*}{\centering GIT} &
\multirow{4}{3.7cm}{\raggedright\textbullet\ High $I_{sp}$ and high efficiency.\\ \textbullet\ Most mature EP technology (with Hall thrusters).}
& NEXT-C &
Proto-flight tested, delivered for DART &
4,150 &
0.69 &
- &
\cite{monheiser2021nextc} \\

& & NEXIS &
Laboratory demonstrated. &
7,500--8,100 &
0.78--0.81 &
20 &
\cite{goebel2004discharge,randolph2004overview} \\

& & HiPEP &
Laboratory tested. &
7,500--10,000 &
$\sim$0.75 &
25 &
\cite{foster2004high} \\

& & RIT-22 &
Precursor RITA-10 demonstrated for ARTEMIS (2002). &
3,000--6,000 &
$>$ 0.8 &
5 &
\cite{loeb2011interstellar,konstantinov2012rit22,killinger2005rita} \\
\midrule

% ----------------------- AF-MPD -----------------------
% The characteristics bullets sit in a normal (auto-growing) p-cell rather than a
% \multirow, so the row expands to fit them and they can never spill past the bottom
% rule (the previous \multirow{4} spanned only two thruster rows -- too short for the
% three bullets). \newline breaks lines inside the cell; the type label spans both rows.
\multirow{2}{*}{\centering AF-MPD} &
{\textbullet\ High $I_{sp}$ and thrust density \cite{boxberger2019current} \newline
\textbullet\ No flight heritage at mission-relevant power; an early pulsed unit was flight-tested \cite{kuriki1981flight}. Laboratory-tested at the 100~kW class \cite{boxberger2017integral,boxberger2019current,sperber2025performance} \newline
\textbullet\ High-power but low TRL} &
SX3 &
100 kW-class lab thruster &
4,665 &
0.62 &
100 &
\cite{boxberger2019current} \\
& & \makecell[l]{CAST/\\AF-MPDT} &
\makecell[l]{150 kW superconducting\\lab thruster} &
5,714 &
0.77 &
150 &
\cite{zheng2021integrated} \\
\bottomrule
\end{tabular}}
\end{widetable}

GITs offer substantial flight heritage and remain the most mature route to high-$I_{\mathrm{sp}}$ operation at high efficiency. AF-MPD thrusters are at lower TRL and have no flight heritage at mission-relevant power. However, the laboratory results in Table~\ref{tab:thrusters_big} approach the reference operating point. If this performance can be sustained with demonstrated lifetime and system-level integration, AF-MPD concepts are credible candidates for higher-power stages, offering high thrust density, compact packaging, and good throttling capability. Flight-proven GITs serve as the conservative baseline. AF-MPDs offer a higher-thrust-density alternative, so fewer units are needed and the clustering complexity of megawatt-class architectures is reduced. In the present model, the EP system is represented as an aggregated thruster cluster with constant $I_{\mathrm{sp}}$ and $\eta$.

%=======================================================================
\section{Reference Mission Definition}
\label{sec:mission}

Building on the preceding review and the HIHT photovoltaic assumptions, we define a reference SEP--SOM mission to quantify delivered payload mass at 200~AU within 25~years.

This objective is aligned with outer-heliosphere concepts such as NASA’s Interstellar Probe concept \cite{brandt2022interstellar,2022mcnutt}, which target heliopause-class distances on multidecadal timescales with a substantial payload to characterise heliospheric structure and its interaction with the very local interstellar medium, including energetic particles, dust, fields, and interstellar plasma. Representative payload elements include plasma and energetic neutral atom sensors, magnetometers, cosmic-ray and dust analysers, UV/visible imagers, and radio and plasma-wave instruments.

While trajectory concepts optimised for ultra-light payload masses could in principle reach $200\,\mathrm{AU}$ in shorter times, the present study adopts a different figure of merit: for a fixed target distance and transfer time, we maximise delivered payload mass. In the adopted lumped-parameter spacecraft model, payload mass is the adjustable part of the dry-mass budget and therefore directly measures allocatable science capability and margin. This choice also supports the sensitivity analysis: it makes the dependence on the enabling parameter $\alpha_{\mathrm{EPS}}$ explicit, with additional sensitivity to the lumped structural/bus mass fraction $\mu_s$ quantified in Section~\ref{subsec:sensitivity_analysis}. From this we can derive the minimum $\alpha_{\mathrm{EPS}}$ at which the architecture delivers positive payload ($m_{pl}>0$).

The mission comprises launch, a SEP spiral that raises aphelion and lowers perihelion to $r_{\mathrm{SOM}}$, the near-perihelion SEP--SOM thrust arc, an optional Jupiter transfer and gravity assist, and a ballistic cruise to 200~AU. Beyond Jupiter orbit, the propulsion system is inactive. The quantitative reference-case choices (baseline $r_{\mathrm{SOM}}$, launcher interface, and whether a JGA is used) are collected in Section~\ref{subsec:scope_limits}.

%=======================================================================

\section{Methodology}
\label{sec:methodology}

Identifying an effective steering strategy for a low-thrust mission is challenging, because at every time step the controller must decide whether to thrust, and with what magnitude and direction. Classical low-thrust trajectory optimisation is typically addressed with indirect optimal-control methods based on Pontryagin's Maximum Principle or with direct transcription and collocation schemes that discretise the trajectory and solve the resulting nonlinear programme. These approaches often rely on adequate initial guesses and expert knowledge in astrodynamics and control theory \cite{Dachwald2019,ohndorf2016multiphase}, and the obtained solutions tend to remain near the initial guess, which does not necessarily coincide with the (unknown) global optimum \cite{dachwald2005optimization}.

Dachwald introduced evolutionary neurocontrol (ENC), in which an evolutionary algorithm optimises the weights of a neural-network controller that maps spacecraft state variables to thrust commands \cite{dachwald2004evolutionary}. The InTrance framework developed by Dachwald and later extended by Ohndorf is the most widely used ENC implementation for low-thrust trajectory design and can employ up to 28 neurocontroller input variables \cite{dachwald2004evolutionary,ohndorf2016multiphase}. Whether a substantially smaller input set is sufficient for the planar SEP--SOM problem was an open question at the outset of this work. We therefore developed the SOMBRERO algorithm in Python: a planar evolutionary neurocontroller with only four spacecraft state inputs and a mutation-based evolutionary operator. This approach avoids the need for carefully crafted initial guesses and provides a global search mechanism suitable for the strongly nonlinear, nonconvex SEP--SOM design space.

In the following, we define the spacecraft model, scope and validity bounds as well as the numerical trajectory integration setup, and the details of the evolutionary scheme used in SOMBRERO.

\subsection{Spacecraft Model}\label{subsec:spacecraft_model}

This study is a conceptual mission-architecture assessment for a solar electric propulsion Oberth maneuver with small perihelia, not a fully detailed mission design. Accordingly, a simplified spacecraft model is employed and explicit mass margins are not applied at this stage. The total spacecraft launch mass $m_0$ is assumed to be composed of four major components:

\begin{itemize}
  \item Mission payload (allocatable dry) mass $m_{pl}$: mass available for the science payload and payload-specific hardware (equivalently, the margin that could instead absorb heavier or lower-performing subsystems than assumed);
  \item Propellant mass $m_{p}$;
  \item Electrical power system mass $m_{EPS}$ (photovoltaic arrays + power processing), which determines the propulsion-relevant solar electric power at 1~AU via $P_{1\,\mathrm{AU}}=m_{EPS}\alpha_{EPS}$;
   \item Structure mass $m_s$, which comprises the thruster cluster and propellant tankage, auxiliary subsystems (e.g., communications and attitude determination and control system), and any deep-space bus-power provision (e.g., RTG). This lumped term does not enter the solar power scaling.
\end{itemize}
For clarity, the delivered dry spacecraft mass after completion of thrusting is $m_{\mathrm{dry}} = m_{pl}+m_s+m_{\mathrm{EPS}}$ (excluding propellant). In this work, the optimisation objective is $m_{pl}$. The corresponding non-dimensional mass fraction with respect to the launch mass is given by $\mu_i = m_i/m_0$. The combined mass fraction of payload and propellant at launch is $\mu_{pp,0} = \mu_{pl}+\mu_{p,0}$.

\subsection{Scope, assumptions, and validity bounds}
\label{subsec:scope_limits}
This paper is an architecture-level feasibility analysis. To avoid repeating caveats across sections, the reference-case inputs and the main validity bounds of the model are summarised here. Sensitivity to key parameters is quantified in Section~\ref{subsec:sensitivity_analysis}.

\subsubsection{Reference-case definition}
The reference case is defined by the following inputs:

\begin{itemize}
\item \emph{Perihelion distance:} The baseline architecture assumes a single SEP--SOM at $r_{\mathrm{SOM}} = 0.3\,\mathrm{AU}$. This choice is motivated by the radiative-equilibrium scaling $r_{\mathrm{SOM}} \propto T^{-2}$ for a temperature-limited array \cite{landis2008solar}. Depending on whether radiative emission is modelled from one side or from both sides of the array, this scaling implies perihelia of order $0.24$--$0.34\,\mathrm{AU}$ for $T\approx400^\circ\mathrm{C}$ \cite{genovese2023advanced} (see~\ref{app:rsom_scaling}). We use $0.3\,\mathrm{AU}$ as a representative central case.
 This Mercury-like perihelion keeps the SEP--SOM well outside the few–$R_\odot$ regime of chemical and solar-thermal SOMs. The thermal environment is then comparable in order of magnitude to that of Mercury orbiters such as MESSENGER.
  \item \emph{Trajectory class:} Two variants are evaluated: a direct trajectory and a Jupiter-gravity-assist (JGA) trajectory. Unless noted otherwise, the JGA case serves as the performance-oriented reference under favourable planetary phasing, and the direct case provides a window-independent alternative.
  \item \emph{Launch interface:} The launch-to-escape interface is represented via a fitted injected-mass model for the fully expendable Falcon Heavy (\ref{app:launch}); after an exploratory phase, $C_3$ and launch angle are fixed to improve convergence (Validity bounds below).

  \item \emph{Electrical power system:} For the reference case, a system-level value of $\alpha_{\mathrm{EPS}}=200\ \mathrm{W\,kg^{-1}}$ at 1~AU is assumed for the overall electrical power system (including solar arrays and power-processing units), see Subsection~\ref{subsub:specific_power_literature_review}; the sensitivity analysis then varies $\alpha_{\mathrm{EPS}}$ to map payload-feasibility thresholds. The HIHT array structures are assumed to be operable at 400$^\circ$C at the perihelion conditions of the reference SEP--SOM (see Subsection~\ref{subsub:HIHT_literature_review}). These are treated as technology assumptions for the reference-case feasibility analysis (i.e., not assumed to be flight-demonstrated).

  \item \emph{Available solar electric power scaling:} In the trajectory optimisation, the available EPS power is modelled as $P_{\max}(r)=P_{1\,\mathrm{AU}}(r/1\,\mathrm{AU})^{-\kappa}$ with an effective exponent $\kappa=1.5$ (instead of the geometric inverse-square law $\kappa=2$) to approximate temperature-related photovoltaic efficiency losses near the Sun \cite{ohndorf2016multiphase,fortescue2011spacecraft}. $\kappa$ is used here as an effective parameter rather than a first-principles thermal model. At $0.3\,\mathrm{AU}$ this reduces the power increase from $(1/0.3)^2\approx11.1$ to $(1/0.3)^{1.5}\approx6.1$, which deliberately avoids over-crediting near-Sun thrust authority in the Oberth-leverage regime (Subsection~\ref{subsec:solar_oberth_theory}). This scaling should be read as a compact approximation of usable electrical power in the inner Solar System, not as a detailed array model across all heliocentric distances.
  \item \emph{Structural mass fraction:} Following the reference heliopause-probe mass budget in \cite{loeb2011interstellar}, the structural (bus and auxiliary subsystems) launch-mass fraction is set to $\mu_s=0.30$. In the \ref{app:validation_InTrance} literature-reproduction case, $\mu_s=0.289$. A comparable value is implied as well by the Mercury-orbiter MESSENGER dry-mass breakdown, $\mu_{s}\approx0.31$ \cite{WertzEverettPuschell2011NewSMAD} (see \ref{app:messenger_mus}).   During the near-perihelion arc, the spacecraft bus is assumed to be protected by a passive, highly reflective sunshade, as on MESSENGER \cite{messenger2004presskit}. There, the Sun-facing surface was predicted to reach $\sim370\,^{\circ}\mathrm{C}$ while the bus behind the sunshade was designed to operate near room temperature ($\sim20\,^{\circ}\mathrm{C}$) \cite{messenger2004presskit}. Hence, no dedicated Parker Solar Probe--style heat-shield mass (or explicit active-cooling mass) is modelled. For a MESSENGER-class mass split, the thermal subsystem fraction is below 5\% of launch mass \cite{WertzEverettPuschell2011NewSMAD} and is therefore subsumed in $m_s$ (and thus in $\mu_s$). Sensitivity to $\mu_s$ is assessed in Section~\ref{subsec:sensitivity_analysis}.

  \item \emph{High-power electric propulsion:} The propulsion architecture is modelled as a modular cluster of high-power electric thrusters (e.g., AF-MPD or gridded ion units in the 50--150~kW class). To accommodate the large variation in available solar power (ranging from $\sim$330~kW at 1~AU to $\sim$2~MW at 0.3~AU), the system is assumed to employ a switching strategy that activates or deactivates individual thruster strings. This operational concept allows active units to function near their nominal design points, supporting the assumption of a constant system-level specific impulse ($I_{sp}=6{,}000$~s) and thrust efficiency ($\eta=0.75$) across the broad power envelope. Detailed throttling maps, duty-cycle limits and lifetime constraints are not explicitly modelled. See the validity bounds below.
\end{itemize}

\subsubsection{Validity bounds}
The results should be interpreted as architecture-level performance potential rather than a fully implementable mission design. The main bounds are:

\begin{itemize}
  \item \emph{Technology targets:} The dominant technology uncertainty in the present study is how quickly high-temperature, high-specific-power solar arrays can be matured from demonstrated cells (Table~\ref{tab:hiht_cells}) to integrated panels and subsystems. In the trajectory and mass model, the electrical power system is represented by a single parameter pair $(T_{\mathrm{max}}, \alpha_{\mathrm{EPS}})$, with HIHT arrays parameterised by $T_{\mathrm{max}} = 400\,^\circ\mathrm{C}$ at perihelion and a constant reference $\alpha_{\mathrm{EPS}} = 200\,\mathrm{W\,kg^{-1}}$ at 1~AU. The pair $(T_{\mathrm{max}},\alpha_{\mathrm{EPS}})$ is treated as independent. In reality the two are coupled: thermal hardening such as radiators and concentrator optics draws on the same mass budget as the array, so reaching both targets together is more demanding than reaching either alone. Because the sensitivity map already spans lower $\alpha_{\mathrm{EPS}}$ (Fig.~\ref{fig:payload_maps}), it indicates how the payload responds if hardening lowers the specific power. This abstracts away panel-level effects such as radiator and harness penalties, temporal degradation, and mechanical and thermal integration with concentration optics and the spacecraft bus. Accordingly, $\alpha_{\mathrm{EPS}}$ is used here as a system-level technology target motivated by Section~\ref{subsec:EPS_literature_review}, not as a flight-demonstrated capability. Section~\ref{subsec:sensitivity_analysis} quantifies the payload penalty for lower system-level EPS specific power. These demands are not unique to the solar-electric route: nuclear-electric propulsion places comparable weight on system-level specific power, and high-performance solar- and laser-sail concepts face first-of-a-kind engineering challenges of their own. Cell-level capability at $400\,^\circ\mathrm{C}$ is already shown in the laboratory, and deployable-array specific power has risen across successive designs. Panel- and subsystem-level integration at that temperature, lifetime testing, and the associated engineering work remain open.

  \item \emph{Spacecraft and subsystem model:} A fixed EPS specific power and structural mass fraction are assumed, together with a generic high-power EP cluster with constant $I_{\mathrm{sp}}$ and efficiency. Ageing, detailed thermal limits, radiation degradation and explicit mass margins are neglected, so the reported payload masses are idealised architecture-level values rather than fully margined system designs.

  \item \emph{Dynamics and operations:} Trajectories are propagated in a planar, two-body (Sun--spacecraft) model. The Jupiter gravity assist is treated analytically, and thrust is assumed to be available whenever commanded by the neurocontroller. Planetary ephemerides, out-of-plane motion, perturbations, duty-cycle limits and lifetime constraints are not explicitly modelled, which may affect achievable $\Delta v$ and thrusting durations in a detailed design.

  \item \emph{Optimisation scope:} The evolutionary neurocontrol framework delivers high-quality, but not provably optimal, steering laws within a restricted design space.

  \begin{itemize}
    \item \emph{Near-optimal solutions:} SOMBRERO is a stochastic optimiser with finite population. The reported trajectories should be regarded as near-optimal for the chosen setup. Additional restarts or hybrid methods could yield modest payload improvements.
  
    \item \emph{Restricted design space:} The optimisation is confined to a fixed architecture (single SOM at $r_{\mathrm{SOM}} \approx 0.3\,\mathrm{AU}$, planar dynamics and a simplified launcher model). After an exploratory phase, $C_3$ and launch angle are fixed to improve convergence, excluding potentially competitive solutions with non-zero $C_3$, alternative perihelia, additional assists or out-of-plane maneuvers.
  
    \item \emph{Controller and objective:} The neurocontroller has a deliberately simple structure with a bounded thrust-angle representation and a single optimisation goal (“maximise payload, provided the 25-year limit is satisfied”). This setup does not map the full trade-off between payload and flight time and may miss more complex thrusting strategies that are locally optimal within a richer control parametrisation.

  \end{itemize}

  The reported payloads are therefore near-optimal realisations of the proposed SEP--SOM architecture, not strict upper bounds. More exhaustive optimisation is expected to refine numerical values, not alter the qualitative conclusions.

\end{itemize}

\subsection{Evolutionary Neurocontrol Framework}\label{subsec:neurocontrol}

We seek a steering policy $P(\boldsymbol{\chi})$ that maps the spacecraft state $\boldsymbol{\chi}$ to controls $\boldsymbol{u}$ and maximises delivered payload mass $m_{pl}$ at $R=200$~AU under a 25-year limit, with initial conditions $\boldsymbol{c_0}$ optimised jointly. The controller genome $\boldsymbol{\xi}$ (weights and biases) is evolved together with $\boldsymbol{c_0}$ as a combined chromosome $\boldsymbol{\pi}=\{\boldsymbol{c_0},\boldsymbol{\xi}\}$.

\subsubsection{Neurocontroller}

The Neurocontroller (NC) is designed to map the transformed current state of the spacecraft, represented as $\boldsymbol{\chi} = \{r,~\nu_{acc},~|\mathbf{v}|,~\varphi\}$, to steering commands $\boldsymbol{u} = \{\theta_{T},~u_{P}\}$, as shown in Fig.~\ref{fig:neurocontroller}.

\makeatletter
\if@twocolumn
  \begin{figure*}[ht]
    \centering
    \includegraphics[width=\textwidth]{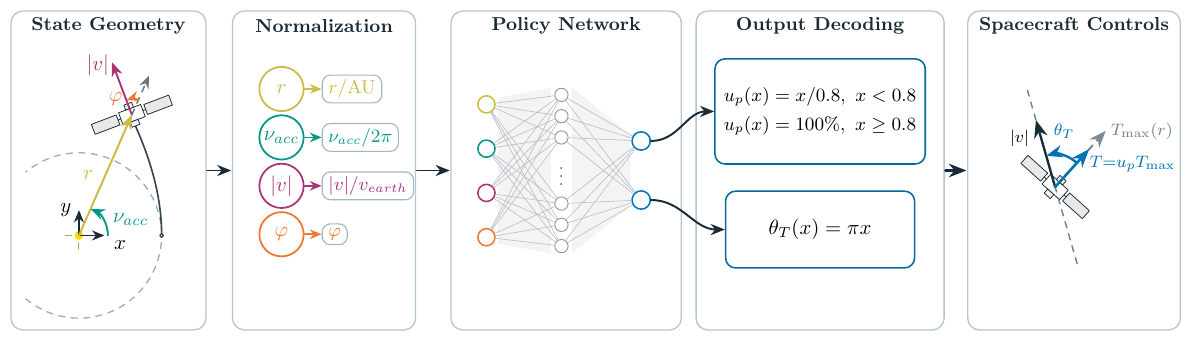}
    \caption{\textbf{Evolutionary neurocontroller used in SOMBRERO.} A feed-forward neural network with one hidden layer and 20 neurons maps the normalised heliocentric state $(r,\nu_{\mathrm{acc}},|\mathbf{v}|,\varphi)$ to thrust angle $\theta_T$ and throttle command $u_P$. Weights and biases encode the steering policy and are optimised with an evolutionary algorithm.}
    \label{fig:neurocontroller}
  \end{figure*}
\else
  \begin{figure}[ht]
    \centering
    \includegraphics[width=\linewidth]{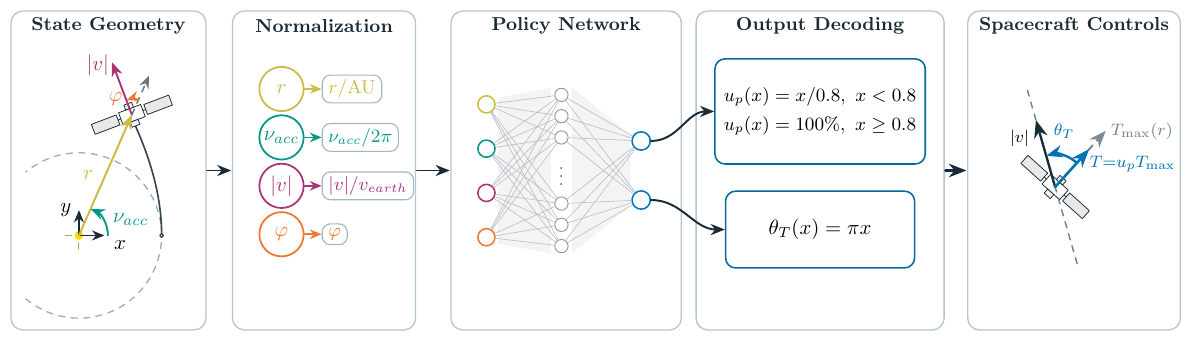}
    \caption{\textbf{Evolutionary neurocontroller used in SOMBRERO.} A feed-forward neural network with one hidden layer and 20 neurons maps the normalised heliocentric state $(r,\nu_{\mathrm{acc}},|\mathbf{v}|,\varphi)$ to thrust angle $\theta_T$ and throttle command $u_P$. Weights and biases encode the steering policy and are optimised with an evolutionary algorithm.}
    \label{fig:neurocontroller}
  \end{figure}
\fi

\makeatother

Here, the normalised heliocentric state is parameterised as:

\begin{itemize}
  \item $r$ is the heliocentric radial distance, normalised to astronomical units (AU),
  \item $\nu_{acc}=(\nu+2\pi N_{rev})/2\pi$ is the accumulated true anomaly, where $N_{rev}$ is the number of completed heliocentric revolutions,
  \item $|\mathbf{v}|$ is the magnitude of the heliocentric velocity, normalised in units of Earth's heliocentric velocity,
  \item $\varphi$ is the angle between the spacecraft's velocity vector and its position vector
\end{itemize}

Together, these inputs summarise the planar heliocentric state relevant to steering. All inputs are normalised to the order of unity to aid learning. The accumulated true anomaly is used to prevent a singularity jump from $2\pi$ to $0$ near the SOM perihelion, which would otherwise obstruct learning. This representation also provides mission time information. 

The outputs of the Neurocontroller are:
\begin{itemize}
  \item $\theta_{T} \in [-\pi, 0]$: the thrust angle with respect to the current direction of flight,
  \item $u_{P} \in [0,1]$: a dimensionless throttle command scaling the electric power (realised via different mass flow rates) supplied to the thrusters, i.e.\ $P(t) = u_{P}(t)\,P_{\max}(r(t))$. We model $u_P$ as a continuous throttle variable representing the aggregate power fraction of a modular thruster cluster. In flight hardware, this profile would be discretised by switching individual thruster strings and modulating the duty cycle of active units. 
\end{itemize}

Bounded activation functions are used to enforce physically meaningful control outputs. The exact functional forms are provided in \ref{app:enc_details}.

\subsubsection{Tournament Selection}

We use binary tournament selection following Dachwald~\cite{dachwald2004evolutionary} and Ohndorf~\cite{ohndorf2016multiphase}. Two individuals are drawn at random. The fitter individual is copied to the next generation. The other is replaced by crossover followed by mutation. We employ a winner--loser crossover, which differs from the winner--winner scheme in InTrance~\cite{ohndorf2011flight}. Exploration is primarily mutation-driven in the present setup.

\subsubsection{Crossover}
We employ four crossover operators (one-point, uniform, node, and arithmetic) with equal probability (25\% each) following Dachwald~\cite{dachwald2004evolutionary} and Ohndorf~\cite{ohndorf2016multiphase}. Operator definitions are provided in \ref{app:enc_details}.

\subsubsection{Mutation}

After crossover, each allele $\pi_j \in \boldsymbol{\pi}$ has a predefined probability $p_m$ of being mutated by a factor sampled from a normal distribution $\mathcal{N}(1, \sigma_m)$, where $\sigma_m$ is the standard deviation of the mutation. Initial conditions are more sensitive to change, thus they are mutated with $\sigma_m/10$.

\subsubsection{Staged Optimisation and Fixed Launch Parameters}\label{subsubsec:opt_schedule}
The SOMBRERO runs reported here use two successive optimisation stages: (i) exploration with \(\sigma_m=0.1\) and \(p_m=0.1\), and (ii) exploitation initialised from the best exploratory individual with reduced mutation probability \(p_m=0.05\).

For the optimised trajectories from Section~\ref{subsec:optimized_trajectories}, the optimiser tended toward \(\theta_{\mathrm{launch}} \rightarrow 0^\circ\) and \(C_3 \rightarrow 0~\mathrm{km^2\,s^{-2}}\) across repeated exploratory runs. For computational efficiency, we therefore fix \(\theta_{\mathrm{launch}}=0^\circ\) and \(C_3=0\) in the reported runs, consistent with the restricted design space stated in Subsection~\ref{subsec:scope_limits}. We hypothesise that this tendency reflects a mass-efficiency trade: energy supplied at launch (\(C_3\)) trades steeply against injected mass, whereas the high-\(I_{sp}\) SEP stage can add the same specific orbital energy at a much smaller mass penalty, so for a payload-maximising objective it is preferable to inject the maximum mass at \(C_3=0\) and add the energy on-board, where the near-Sun Oberth pass offers the greatest leverage.

\subsubsection{Fitness Functions}
The optimisation is guided by a staged fitness score $J = \sum_{i=1}^{5} j_i$ that enforces the mission sequence before activating the payload objective. The five stages are:

\begin{enumerate}
  \item \textbf{Solar Oberth Maneuver:} Evaluates the spacecraft's ability to achieve the precise design perihelion ($r_{SOM,design}$) required for the SOM, penalising deviations.
  \item \textbf{Approach Jupiter:} Rewards the trajectory for reaching Jupiter's orbital radius (5.2 AU) post-SOM.
  \item \textbf{Reaching Solar Escape Velocity:} Rewards the kinetic energy after the Jupiter Gravity Assist (JGA) needed to reach solar escape velocity.
  \item \textbf{Time to 200 AU:} Penalises solutions that fail to reach the 200 AU target within the 25-year mission constraint.
  \item \textbf{Maximise Payload Mass:} The final optimisation objective, activated only after all preceding constraints are satisfied. This score is directly proportional to the deliverable payload mass ($m_{pl}$).
\end{enumerate}

This successive evolutionary staging was effective for the present optimisation setup. The specific mathematical formulations for each fitness function, along with the criteria for the initial heuristic filter, are detailed in \ref{app:fitness}.

\subsection{Simulation Setup}
A heliocentric two-body problem low-thrust trajectory simulation environment with dynamic time steps is developed in Python to evaluate trajectories from various steering policies. The initial conditions $\mathbf{c_0}$ are: launch angle $\theta_{launch}$, hyperbolic excess energy $C_3$ and combined mass fraction of propellant and payload at launch $\mu_{pp,0}$.

\subsubsection{Launch}

\begin{figure}[h!]
 \centering
 \includegraphics[width=\linewidth]{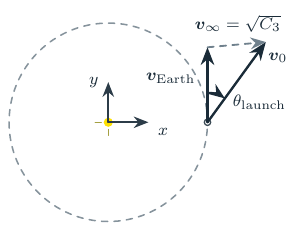}
 \caption{\textbf{Launch parametrisation used in the planar trajectory model.} The launch angle $\theta_\mathrm{launch}$ and injected energy $C_3$ define the initial injection state.}
 \label{fig:launch_control}
\end{figure}

The spacecraft is launched from Earth with an initial mass $m_0$ and a launch angle $\theta_{launch}$ (counterclockwise positive) with respect to Earth's heliocentric velocity, as shown in Fig.~\ref{fig:launch_control}.

Considering the Earth-relative hyperbolic excess velocity from the launcher $v_{\infty,\text{Earth}} = \sqrt{C_3}$, the initial velocity is given by:

\begin{equation}
  \mathbf{v}_0 = \begin{bmatrix}
-\sin(\theta_{\text{launch}})\, v_{\infty,\text{Earth}}, & \cos(\theta_{\text{launch}})\, v_{\infty,\text{Earth}} + v_{\text{Earth}}
\end{bmatrix}^T
\end{equation}

C$_3$ determines the initial launch mass\cite{ohndorf2011flight}:

\begin{equation}
  m_0=m_{0,max} \exp(-\gamma \cdot C_3)
\end{equation}

$m_{0,max} = 15{,}189$~kg and decay parameter $\gamma=0.02165\,\mathrm{s^2\,km^{-2}}$ are derived by fitting conservative performance curves for the fully expendable Falcon Heavy vehicle \cite{nasa_elv_performance}, as shown in \ref{app:launch}. This model yields slightly lower injected mass estimates than some publicly reported manufacturer performance figures (e.g., $16.8$~t to Mars transfer \cite{spacex_falcon_heavy}), thereby introducing an implicit mass margin at the launch interface of at least 1.6~t.

\subsubsection{Trajectory Propagation}

After launch, the spacecraft acceleration $\mathbf{a_{sc}} = \begin{bmatrix} \ddot{x} & \ddot{y} \end{bmatrix}^T$ is given by the superposition of the sun's gravitational acceleration $\mathbf{a_G}$ and thrust acceleration $\mathbf{F_T}/m_{sc}$. Spacecraft steering is achieved through control of the thrust angle $\theta_T$ and the power throttling $u_P$, as shown in Fig.~\ref{fig:neurocontroller}.

\begin{equation}
\mathbf{a}_{sc} = \begin{bmatrix}\ddot{x}\\ \ddot{y}\end{bmatrix} = \mathbf{a}_G + \frac{\mathbf{F_T}}{m_{sc}} 
= -\begin{bmatrix}x\\ y\end{bmatrix}\frac{GM_\odot}{\sqrt{x^2+y^2}^3}
+ \left(\mathbf{n_T}(\theta_T)\frac{2\eta\,u_P P_{\max}(r)}{g_0I_{sp}}\right)/m_{sc}
\end{equation}

where $x$ and $y$ are the heliocentric coordinates, $GM_\odot$ is the standard gravitational parameter of the Sun, $\mathbf{n_T}(\theta_T)$ is the thrust direction unit vector as a function of the thrust angle $\theta_T$, $\eta$ is the thrust efficiency, $P$ is the power, $u_P$ is the power throttle, $g_0$ is the standard gravitational acceleration, and $I_{sp}$ is the specific impulse. The trajectory propagation is carried out using an adaptive Runge--Kutta (RK45, Dormand--Prince) scheme with error-controlled step size.

\subsubsection{Jupiter Gravity Assist}\label{subsec:jupiter_gravity_assist}

When the JGA option is enabled, we apply a planar, instantaneous flyby model in which the post-encounter heliocentric velocity is obtained by rotating the Jupiter-relative velocity vector. Equations and parameter definitions are given in \ref{app:jga_model}.

%=======================================================================
\section{Results}
\label{sec:results}

All results in this section are generated within an idealised planar heliocentric two-body model with no out-of-plane dynamics or ephemeris-based multi-body effects. When included, the JGA follows the analytical in-plane velocity-rotation model of Section~\ref{subsec:jupiter_gravity_assist}. Accordingly, the results should be interpreted as architecture-level performance potential rather than a fully implementable mission design. More exhaustive 3D, ephemeris-based optimisation is expected to refine timing and numerical payload values, not to alter the qualitative trade-offs reported here. Model verification and validation are reported in Sections~\ref{sec:verification} and \ref{sec:validation}. The SOMBRERO optimisation protocol (including staged exploration/exploitation and fixed launch parameters) is summarised in Section~\ref{subsubsec:opt_schedule}.

\subsection{Validation of the SOMBRERO Model} \label{sec:validation}

Validation is performed by reproducing the reference SOM+JGA SEP stage from \cite{loeb2011interstellar,ohndorf2011flight} using the same initial conditions (see ~\ref{app:trajectory_validation}). The subsequent RTG-powered segment used in \cite{ohndorf2011flight} to reduce $t_{200\,\mathrm{AU}}$ is not modelled here. The obtained results are summarised in Table~\ref{tab:comparison_validation_results}.

\begin{table}[htb]
  \centering
 \caption{\textbf{Comparison of InTrance and SOMBRERO for the SOM with Jupiter gravity assist (SOM+JGA) case at $r_{\mathrm{SOM}}=0.7~\mathrm{AU}$.} Masses refer to the SEP stage. The literature $t_{200\,\mathrm{AU}}$ includes the additional RTG-powered segment.}
  \begin{tabular}{cccc}
    \hline
    \textbf{Quantity} & \textbf{Literature} & \textbf{This Work} & \textbf{Difference (\%)} \\    
    \hline
    $v_{\text{post-JGA}} \mathrm{[km\,s^{-1}]}$  & 39.3  & 38.8  & -1.3 \\
    $\Delta v_{\text{JGA}}$ [$\mathrm{km\,s^{-1}}$] & 12.5  & 12.3  & -1.6 \\
    $m_{\text{pl}}$ [kg]     & 498 & 464 & -6.8 \\
    $m_{\text{EPS}}$ [kg]    & 265 & 266 & 0.4  \\
    $m_{\text{p}}$ [kg]     & 440 & 478 & 8.6  \\
    $m_{\text{s}}$ [kg]     & 489 & 491 & 0.4  \\
    $m_{0}$ [kg]        & 1692 & 1700 & 0.5  \\
    $t_{\text{200AU}}$ [years]  & 23.8  & 30.4 & 27.7 \\
    \hline
  \end{tabular}
  \label{tab:comparison_validation_results}
\end{table}

Differences are in general small and can be explained by optimisation towards slightly different aphelions. Together with the sub-$1\ \mathrm{m\,s^{-1}}$ thrust-integration error reported below, this supports the numerical accuracy of the trajectory propagation (see \ref{app:trajectory_validation}).

The only notable discrepancy is $t_{200\,\mathrm{AU}}$, which is not directly comparable here: the reference value includes a subsequent RTG-powered post-SEP stage, whereas the rebuilt case reports the SEP-stage solution and the ensuing ballistic coast. The resulting $\Delta t_{200\,\mathrm{AU}} \approx 6.6$~yr is therefore expected and consistent with the $\sim$6~yr reduction attributed to the added RTG stage in~\cite{loeb2011interstellar}. The close agreement in SEP-stage mass and post-JGA velocity in Table~\ref{tab:comparison_validation_results} supports the accuracy of the core propulsive integration.

\subsection{Optimised Trajectories and Mission Profiles}\label{subsec:optimized_trajectories}

The following subsections detail the trajectory characteristics, steering laws, and mass budgets of the optimised solutions. We first present the mission profile incorporating a Jupiter Gravity Assist (JGA), which yields higher payload but is sensitive to Earth-Jupiter-heliopause phasing and entails additional operational constraints, followed by the direct trajectory scenario, which avoids those constraints at the cost of reduced payload. Based on preliminary exploratory runs, \(\theta_{\mathrm{launch}}\) and \(C_3\) are fixed to \(0^\circ\) and \(0~\mathrm{km^2\,s^{-2}}\), respectively, as described in Subsection~\ref{subsubsec:opt_schedule}.

\subsubsection{Results Incorporating Jupiter Gravity Assist}
\begin{figure*}[t!]
  \centering
  \includegraphics[width=\textwidth]{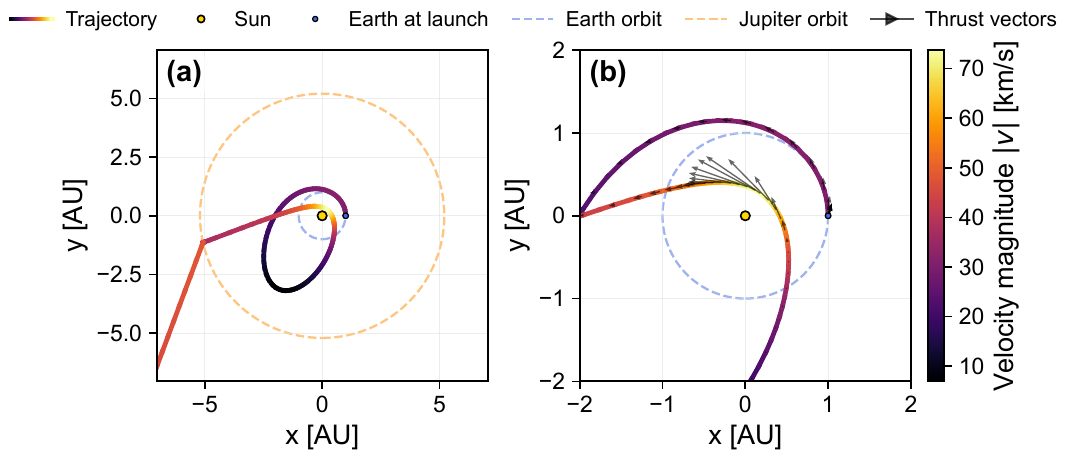}
  \caption{\textbf{Optimised SEP--SOM trajectory with Jupiter gravity assist.}
  Trajectory in the heliocentric in-plane frame; marker colour encodes the heliocentric \emph{velocity magnitude} $|v|$ (colourbar).
  Dashed circles indicate the Earth and Jupiter orbits; the blue marker indicates Earth at launch and the yellow disk denotes the Sun.
  (a) Full in-plane trajectory, including the Jupiter encounter. (b) Zoomed view of the near-Sun arc. SEP, solar electric propulsion; SOM, solar Oberth maneuver; JGA, Jupiter gravity assist.}
  \label{fig:trajectory_results_with_JGA}
\end{figure*}

The optimised SEP--SOM trajectory including JGA delivers $3{,}083\,\mathrm{kg}$ payload to $200\,\mathrm{AU}$ in $24.97\,\mathrm{years}$ (Fig.~\ref{fig:trajectory_results_with_JGA}). Table~\ref{tab:SOM_JGA_results_summary} summarises the resulting mass and performance breakdown. This JGA contribution is computed at a fixed periapsis under favourable Earth--Jupiter phasing, so it is an optimistic estimate. Less favourable phasing or approach geometry would yield a smaller assist. 

\begin{table}[htb]
  \centering
  \caption{\textbf{Reference-case results summary for the optimised SEP--SOM trajectory including a Jupiter gravity assist.}}
  \begin{tabular}{lll}
    \hline
    \textbf{Quantity} & \textbf{Symbol} & \textbf{Value} \\
    \hline
    Power at 1 AU & $P_0$ & 334.2 kW \\
    Perihelion radius & $r_\text{SOM}$ & 0.308 AU \\
    Power at $r_\text{SOM}$ & $P_\text{max}$ & 1,954.9 kW \\
    Thrust at $r_\text{SOM}$ & $F_{T,\text{max}}$ & 49.8 N \\
    Velocity change & $\Delta v_\text{EP}$ & $28{,}798\,\mathrm{m\,s^{-1}}$ \\
    Jupiter GA velocity change & $\Delta v_\text{JGA}$ & $10{,}627\,\mathrm{m\,s^{-1}}$ \\
    Total launch mass & $m_0$ & 15{,}189 kg \\
    Payload mass & $m_\text{pl}$ & 3{,}083 kg \\
    Electrical power system mass & $m_\text{EPS}$ & 1{,}671 kg \\
    Propellant mass & $m_\text{p}$ & 5{,}878 kg \\
    Structure mass & $m_\text{s}$ & 4{,}557 kg \\
    Flight time to $200~AU$ & $t_\text{200AU}$ & 24.97 years \\
    Heliocentric excess speed & $v_{\infty,\odot}$ & $9.30\,\mathrm{AU\,year^{-1}}$\\
    \hline
  \end{tabular}
  \label{tab:SOM_JGA_results_summary}
\end{table}

The optimised allocatable payload mass is nearly two orders of magnitude higher than the 35 kg instrument payload found by InTrance for the $r_{SOM}=0.7~\text{AU}$ case. Three factors account for this. i) The reference scenario is launched with Ariane 5 ($C_3=45.1\,\mathrm{km^2\,s^{-2}}$) and thus carries an $m_0$ of only 1692 kg, whereas the assumed launch with an expendable Falcon Heavy $m_0$ is 15{,}189 kg ($C_3=0\,\mathrm{km^2\,s^{-2}}$). This explains one order of magnitude in difference. The residual order of magnitude is attributed to ii) the $r_{SOM}$ at 0.3 AU, which offers a stronger solar Oberth effect. This allows for a higher heliocentric excess speed, which in turn removes the need for a REP stage to satisfy the 25-year time limit. Factor iii) follows: in the optimised case the scientific payload is carried directly on the SEP stage. Previously it had to ride on a further REP stage, whose system mass reduced the deliverable payload.

\subsubsection{Results for Direct Trajectory}
The results of the optimised SOM with no JGA yield a final payload mass of 1,551~kg with 24.34~years of flight time. The trajectory is very similar to the one shown in Fig.~\ref{fig:trajectory_results_with_JGA} and can be found in \ref{app:trajectory_without_JGA}. The results are summarised in Table~\ref{tab:results_summary_no_JGA}.

\begin{table}[htb]
  \centering
  \caption{\textbf{Reference-case results summary for the optimised direct SEP--SOM trajectory without a Jupiter gravity assist.}}
  \begin{tabular}{lll}
    \hline
    \textbf{Quantity} & \textbf{Symbol} & \textbf{Value} \\
    \hline
    Power at 1 AU & $P_0$ & 424.1 kW \\
    Perihelion radius & $r_\text{SOM}$ & 0.303 AU \\
    Power at $r_\text{SOM}$ & $P_\text{max}$ & 2{,}542.8 kW \\
    Thrust at $r_\text{SOM}$ & $F_{T,\text{max}}$ & 64.8 N \\
    Velocity change & $\Delta v_\text{EP}$ & $36{,}191\,\mathrm{m\,s^{-1}}$ \\
    Total launch mass & $m_0$ & 15{,}189 kg \\
    {Payload mass} & ${m_\text{pl}}$ & {1,551 kg} \\
    Electrical power systems mass & $m_\text{EPS}$ & 2{,}120 kg \\
    Propellant mass & $m_\text{p}$ & 6{,}961 kg \\
    Structure mass & $m_\text{s}$ & 4{,}557 kg \\
    {Flight time} & $t_\text{200AU}$ & {24.34 years} \\
    Heliocentric excess speed & $v_{\infty,\odot}$ & $9.03\,\mathrm{AU\,year^{-1}}$\\
    \hline
  \end{tabular}
  \label{tab:results_summary_no_JGA}
\end{table}

For context, we compare the SEP--SOM architecture to a ``vanilla'' outward
electric--propulsion (EP) spiral starting from a circular 1~AU orbit. For the same EP $\Delta v$ as the direct SEP--SOM case (Table~\ref{tab:results_summary_no_JGA}), such a reference spiral departing Earth's orbit ($29.8~\mathrm{km\,s^{-1}}$) would yield a heliocentric hyperbolic excess speed of approximately \mbox{$v_{\infty,\odot} \approx (36.2-29.8)~\mathrm{km\,s^{-1}} \approx 6.4~\mathrm{km\,s^{-1}}$}, corresponding to \mbox{$1.35~\mathrm{AU\,year^{-1}}$}. The associated increase in specific orbital
energy relative to a circular 1~AU orbit is
\begin{equation}
\begin{aligned}
\Delta\varepsilon_\text{spiral}
&= \frac{v_{\infty,\odot}^2}{2} + \frac{GM_\odot}{2a} \\
&= \frac{\bigl(1.35 \times 4,740.47~\mathrm{m\,s^{-1}}\bigr)^2}{2}
 + \frac{1.3271\times 10^{20}}{2 \times 1.49598\times 10^{11}}~\mathrm{m^2\,s^{-2}} \\
&\simeq 4.64\times 10^{8}~\mathrm{m^2\,s^{-2}}.
\end{aligned}
\end{equation}
For the optimised SEP--SOM trajectory, the corresponding specific orbital
energy increment is
\mbox{$\Delta\varepsilon_\text{SEP--SOM} \simeq 1.37\times 10^{9}~
\mathrm{m^2\,s^{-2}}$}, i.e.\ almost a factor of three larger:
\begin{equation}
\frac{\Delta\varepsilon_\text{SEP--SOM}}{\Delta\varepsilon_\text{spiral}}
\approx 2.95 \;\approx\; 3.
\end{equation}
For comparable EP capability, the SEP--SOM architecture
converts onboard $\Delta v$ into heliocentric orbital energy about three times more
efficiently than a conventional EP spiral. This ratio essentially makes the analytic Oberth leverage of Eq.~\eqref{eq:oberth_cont_general} concrete. The spiral is an idealised same-$\Delta v$ reference.

\subsubsection{Verification} \label{sec:verification}

The trajectory-propagation and bookkeeping implementation is verified by comparing (a) the total specific orbital energy change and (b) the EP-induced velocity change, each computed independently via analytical expressions and a discrete numerical approximation for two representative cases (direct and JGA trajectories).

The analytical change in specific orbital energy between the initial circular orbit and Solar System escape is given by
\begin{equation}
  \Delta \varepsilon_{analytic} = \varepsilon_{escape} - \varepsilon_{start}
  = \frac{v_{\infty,\odot}^2}{2} + \frac{GM_\odot}{2\, \cdot 1 \mathrm{AU}}.
\end{equation}
where $v_{\infty,\odot}$ denotes the heliocentric hyperbolic excess speed (distinct from the launcher-provided $v_{\infty,\text{Earth}}=\sqrt{C_3}$).

The specific orbital energy gain from the EP system is evaluated numerically via a discrete approximation
\begin{equation}\label{eq:depsilon_sim_discrete}
   \Delta \varepsilon_{EP,sim} = \int_{t_0}^{t_f} \mathbf{a}_{EP}(t) \cdot \mathbf{v}(t)\, dt\approx \sum_{i=0}^{N} \frac{F_{T,i}}{m_{sc,i}}\, v_i\, \cos\bigl(\theta_{T,i}\bigr) \, \Delta t_i, 
\end{equation}
where $F_{T,i}$ denotes the thrust magnitude at the $i-$th time step. Similarly, the spacecraft velocity change is determined by two independent methods. First, by numerically integrating the acceleration magnitude, given by
\begin{equation}\label{eq:dvsim_discrete}
  \Delta v_{sim} = \int_{t_0}^{t_f} \frac{F_T(t)}{m_{sc}(t)}\, dt \approx \sum_{i=0}^{N} \frac{F_{T,i}}{m_{sc,i}}\, \Delta t_i.
\end{equation}

In contrast, the classical Tsiolkovsky equation provides
\begin{equation}
  \Delta v_{analytic} = g_0 \, I_{sp} \ln \left(\frac{m_{sc,0}}{m_{sc,f}}\right),
\end{equation}
with $g_0 = 9.81\,\mathrm{m\,s^{-2}}$ and an effective specific impulse corresponding to 6,000\,s.

The specific orbital energy gain due to the JGA maneuver is computed by
\begin{equation}
  \Delta \varepsilon_{JGA} = v\,\Delta v_{JGA} + \frac{1}{2}\Delta v_{JGA}^2.
\end{equation}

Table~\ref{tab:propulsion_results} summarises the results for the two cases: one without a JGA maneuver and one including a JGA maneuver.

\begin{table}[htb]
  \centering
  \caption{\textbf{Analytical versus numerical consistency check for $\Delta\varepsilon$ and $\Delta v$ in the direct and Jupiter-gravity-assist cases.}}
  \begin{tabular}{lcc}
    \hline
    \textbf{Quantity} & \textbf{Direct Trajectory} & \shortstack{\textbf{Jupiter}\\ \textbf{Gravity Assist}} \\
    \hline
    $\Delta\varepsilon_{EP,sim}$ [$\mathrm{J\,kg^{-1}}$] & $1.3653\times10^{9}$ & $9.5907\times10^{8}$ \\
    $\Delta\varepsilon_{JGA,sim}$ [$\mathrm{J\,kg^{-1}}$] & $0$ & $4.5486\times10^{8}$ \\
    $\Delta\varepsilon_{sim}$ [$\mathrm{J\,kg^{-1}}$] & $1.3653\times10^{9}$ & $1.4139\times10^{9}$ \\
    $\Delta\varepsilon_{analytic}$ [$\mathrm{J\,kg^{-1}}$] & $1.3597\times10^{9}$ & $1.4144\times10^{9}$ \\
    $\Delta\varepsilon$ discrepancy [\%] & $+0.42$ & $-0.03$ \\
    \hline
    $\Delta v_{sim}$ [$\mathrm{m\,s^{-1}}$] & $36{,}190.73$ & $28{,}798.25$ \\
    $\Delta v_{analytic}$ [$\mathrm{m\,s^{-1}}$] & $36{,}086.73$ & $28{,}770.57$ \\
    $\Delta v$ discrepancy [\%] & $+0.29$ & $+0.10$ \\
    \hline
  \end{tabular}
  \label{tab:propulsion_results}
\end{table}

The sub-percent agreement between analytical and numerical $\Delta\varepsilon$ and $\Delta v$ estimates supports the numerical consistency of the trajectory propagation and EP/JGA bookkeeping for the considered cases. Residual differences are consistent with finite export resolution and numerical quadrature on rapidly varying high-thrust segments near perihelion, rather than indicating a physical inconsistency.

\subsection{Thrust Phasing and Oberth Energy Accumulation}
\label{sec:learned_steering_jga}

\begin{figure}[!t]
  \centering
  \includegraphics[width=\columnwidth]{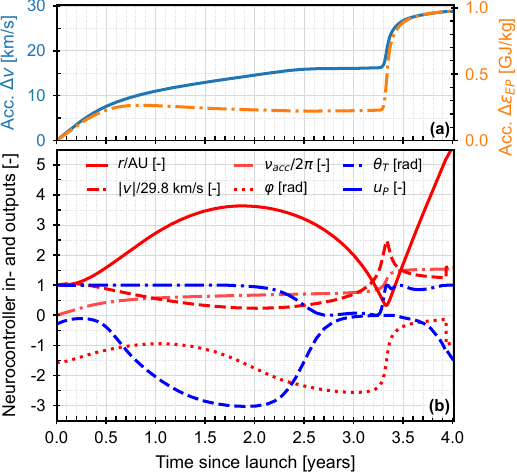}
\caption{\textbf{Accumulated electric-propulsion performance and controller signals versus time for the optimised SEP--SOM trajectory in the Jupiter-gravity-assist case.} Upper panel: cumulative EP $\Delta v$ and EP-induced specific orbital energy gain $\Delta\varepsilon_{\mathrm{EP}}$. Most of $\Delta\varepsilon_{\mathrm{EP}}$ is accumulated during the short near-perihelion thrust arc, although only part of the total $\Delta v$ is delivered there. Lower panel: controller inputs (red) and outputs (blue) including thrust angle $\theta_T$ and throttle $u_P$.}
  \label{fig:neuro_jga_io}
\end{figure}

The upper panel of Fig.~\ref{fig:neuro_jga_io} illustrates how cumulative EP $\Delta v$ and EP-induced specific orbital energy gain $\Delta\varepsilon_{\mathrm{EP}}$ accumulate along the optimised SEP--SOM trajectory in the JGA case. The solution applies $\sim29~\mathrm{km\,s^{-1}}$ of EP $\Delta v$ over the full trajectory, of which $\sim10~\mathrm{km\,s^{-1}}$ are delivered in a short near-perihelion thrust segment. Despite this, the near-perihelion segment produces the dominant increase in EP-induced specific orbital energy gain $\Delta\varepsilon_{\mathrm{EP}}$, consistent with Eq.~\eqref{eq:oberth_cont_general}, which weights the EP energy gain by the instantaneous heliocentric speed.

The lower panel reports the controller outputs, namely thrust angle $\theta_T$ and throttle $u_P$ (blue traces in Fig.~\ref{fig:neuro_jga_io}). During the near-perihelion thrust segment, the controller commands near-prograde thrust angles with throttle close to unity, consistent with concentrating propulsion work near perihelion. A five-phase structure is apparent; \ref{app:learned_strategy_no_JGA} gives the phase breakdown for the no-JGA case, where the structure is similar, and compares the two.

\subsection{Sensitivity Analysis}\label{subsec:sensitivity_analysis}

The presented results generalise beyond a specific mass and power realisation, provided the same reference trajectory is preserved. That requires (a) identical initial conditions and (b) an identical commanded acceleration vector over time. In the following rescaling map, we vary the system-level $\alpha_{\mathrm{EPS}}$ (arrays + PPU, referenced to 1~AU) and $\mu_s$ while holding the thruster model parameters ($I_{sp}=6{,}000$~s, $\eta=0.75$) fixed; the trajectory and propellant fraction are taken from the optimised reference case and are not re-optimised at each grid point. As shown in \ref{app:gen_derivation}, this yields a closed-form relation that maps the discrete simulation reference case $(\alpha_{\mathrm{EPS,sim}},\,\mu_{\mathrm{EPS,sim}},\,\mu_{\mathrm{p}})$ to alternative combinations of $\alpha_{\mathrm{EPS}}$ and $\mu_s$:

\begin{equation}
\begin{aligned}
  m_{\text{pl}} &= f(m_0,\mu_s,\alpha_{\mathrm{EPS}})\\
  &=\;
  m_0\left[\bigl(1 - \mu_{\text{p}}\bigr) - \mu_{\text{s}}
  \;-\; \frac{\alpha_{\mathrm{EPS,sim}}\,\mu_{\mathrm{EPS,sim}}}{\alpha_{\mathrm{EPS}}}\right]
\end{aligned}
\label{eq:mpl}
\end{equation}

\begin{figure*}[h!]
  \centering
  \includegraphics[width=\linewidth]{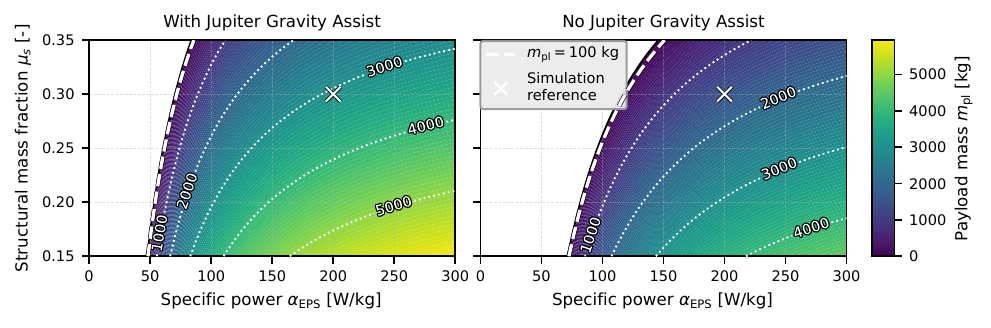}
  \caption{\textbf{Payload mass sensitivity to EPS specific power $\alpha_{\mathrm{EPS}}$ and structural mass fraction $\mu_s$, computed from \eqref{eq:mpl}.} Left: trajectory including a Jupiter Gravity Assist (JGA). Right: direct trajectory without JGA. Colour indicates payload mass $m_{\mathrm{pl}}$ in kg. White dotted isolines are spaced by $1{,}000$~kg and labelled. The thin black curve marks the feasibility boundary $m_{\mathrm{pl}}=0$. The white dashed curve highlights $m_{\mathrm{pl}}=100$~kg. The white $\times$ denotes the simulation reference at $(\alpha_{\mathrm{EPS}},\mu_s)=(200~\mathrm{W\,kg^{-1}},\,0.30)$. At $\mu_s=0.30$, the $100$~kg isoline intersects $\alpha_{\mathrm{EPS}}\approx71.80~\mathrm{W\,kg^{-1}}$ (with JGA) and $\approx118.78~\mathrm{W\,kg^{-1}}$ (without JGA).}
  \label{fig:payload_maps}
\end{figure*}

From Eq.~\eqref{eq:mpl} we can assess the payload impact of EPS specific power $\alpha_{\mathrm{EPS}}$ and structural mass fraction $\mu_s$, holding the reference propellant fraction $\mu_{\mathrm{p}}$, thrust history, and assumed thruster performance fixed. 
Fig.~\ref{fig:payload_maps} shows that, at $\mu_s=0.30$, a $100$~kg payload remains feasible for $\alpha_{\mathrm{EPS}}\approx71.80~\mathrm{W\,kg^{-1}}$ with a JGA, whereas the direct case requires $\approx118.78~\mathrm{W\,kg^{-1}}$ (a $\sim1.7\times$ higher specific-power requirement). Relative to the present-day order-of-magnitude EPS-level value $\alpha_{\mathrm{EPS}}\approx78.8~\mathrm{W\,kg^{-1}}$ from Section~\ref{subsec:EPS_literature_review}, these $100$~kg thresholds correspond to roughly $0.9\times$ (JGA) and $1.5\times$ (direct), under the additional assumption of $400^\circ$C survivability. Ton-class payloads ($\geq 1000$~kg) become feasible at $\alpha_{\mathrm{EPS}}\approx89.02~\mathrm{W\,kg^{-1}}$ with JGA and $\approx158.80~\mathrm{W\,kg^{-1}}$ without, corresponding to about $1.1\times$ and $2.0\times$, respectively, of that same present-day reference.

%=======================================================================
\section{Discussion}\label{sec:discussion}

This section discusses the architecture-level implications and main validity bounds of the results.

The SOMBRERO optimisation results indicate that, under the baseline assumptions (Section~\ref{subsec:scope_limits}), a single SEP--SOM at $r_\mathrm{SOM}\approx0.3\,\mathrm{AU}$ launched on an expendable Falcon Heavy can deliver about $3{,}083\,\mathrm{kg}$ of payload to $200\,\mathrm{AU}$ in $\approx25.0\,\mathrm{yr}$ via a Jupiter Gravity Assist, or $1{,}551\,\mathrm{kg}$ on a direct trajectory in $24.3\,\mathrm{yr}$ (Tables~\ref{tab:SOM_JGA_results_summary} and \ref{tab:results_summary_no_JGA}). These payloads are obtained on a conservatively de-rated Falcon Heavy model, at least 1.6~t below the manufacturer-quoted performance, so they carry an implicit positive launch-interface margin. For the same low-thrust $\Delta v$, the near-Sun thrust arc yields a heliocentric specific-orbital-energy increase $\Delta\varepsilon$ almost three times larger than a $1\,\mathrm{AU}$ outward spiral, and Eq.~\eqref{eq:mpl} and Fig.~\ref{fig:payload_maps} further show that payload feasibility spans from near present-day conventional EPS specific-power levels (e.g.\ $\sim100$~kg with JGA) to about twice those levels for direct ton-class cases, depending on trajectory class and $\mu_s$.

The subsections below focus on the trade-space shift that occurs when SEP is treated as a short-duration, near-Sun high-power phase, instead of a continuously power-constrained cruise architecture whose thrust authority rapidly diminishes with heliocentric distance.

\subsection{High-Temperature Solar Cells as Interstellar Mission Enablers}

High-intensity, high-temperature solar cells have historically been developed for survivability in extreme inner Solar System environments, as reflected in near-Sun mission requirements for high-bandgap devices~\cite{scheiman1999high} and modern spacecraft designs (Parker Solar Probe, MESSENGER, Venus probes)~\cite{Landis2023_chapter, Landis2020_Venus}. This development focus on thermal survival has resulted in limited assessment of HIHT photovoltaics as propulsion-enabling elements, despite the space-photovoltaics community's emphasis on efficiency, specific power, and radiation hardness~\cite{bermudez2021environments, verduci2022solar, li2021brief}. At the same time, trajectory analyses for solar electric Oberth Maneuvers have been limited by the thermal constraints of conventional arrays, restricting prior studies to conservative perihelia of $0.5$–$0.7$\,AU \cite{ohndorf2011flight, bering2021solar, Kluever1997_SEP_Interstellar}. 
 
The present study shows that HIHT photovoltaics can move from a passive survival role to an enabling role for high-power, close-perihelion propulsion architectures for rapid-escape, outer-planet, and interstellar-precursor missions. For example, the 0.3~AU SEP--SOM architecture studied here is, to our knowledge, the first solar-electric Oberth concept proposed at a Mercury-class perihelion ($\sim$0.3~AU), a regime that conventional-array thermal limits had closed to solar-electric propulsion and that the assumed HIHT arrays could open. Prior solar-electric Oberth studies remained at 0.5--0.7~AU (Table~\ref{tab:som_literature}).

\subsection{Solar and Nuclear Electric Propulsion: A New Architectural Perspective on Interstellar Precursors}

For decades, conceptual studies of fast, interstellar precursor missions have favoured nuclear electric propulsion (NEP), widely “advocated as a primary solution’’ \cite{McNutt2025_IAC_NEP_PhysicalLimits} for its distance-independent, constant power \cite{duchek2024nep}, while conventional solar electric propulsion, limited by the inverse-square law, has often been viewed as challenging for outer Solar System missions and considered insufficient for cases like a Pluto orbiter \cite{Woodcock2002NEPOuterPlanets}. While this limitation holds for traditional continuous SEP cruise, using SEP as a short-duration, high-power near-Sun phase can reduce its impact. With HIHT photovoltaics, our model of a $0.3~\mathrm{AU}$ SEP--SOM provides $\sim2~\mathrm{MW}$ near perihelion before coasting, lessening the need for distant solar power. This suggests a possible non-nuclear-propulsion route for high-payload precursor missions when fission-based power is impractical.

Because the SEP--SOM thrust phase spans a significant fraction of the perihelion passage, not a single impulsive burn, it could provide additional
freedom to trim injection dispersions and navigation errors as the spacecraft approaches and recedes from perihelion. This would be expected to mitigate the sensitivity to injection errors that characterises low-perihelion, high-thrust SOM architectures discussed in Section~\ref{subsec:solar_oberth_literature_review}.

\subsection{Implications for Related High-Energy Missions}
\label{sec:implications_related}

The high-payload escape performance shown here (see Tables \ref{tab:SOM_JGA_results_summary} and \ref{tab:results_summary_no_JGA}) also carries over to high-energy missions within the Solar System. A 0.3~AU SEP--SOM can front-load a large increase in specific orbital energy and may reduce transit times for heavy orbiters toward the ice giants relative to multi-gravity-assist architectures, subject to re-optimisation of the outbound leg and arrival conditions. 

For outer-planet orbit insertions (e.g.\ Uranus), a SEP--SOM leg does not inherently impose a large \emph{planetocentric} arrival hyperbolic excess speed $v_{\infty,\text{planet}}$: by re-optimising the inner-leg low-thrust steering, one can target a prescribed $v_{\infty,\text{planet}}$, trading longer flight times for higher delivered payload. Because thrusting is confined to the inner Solar System, the SEP--SOM sets the outbound heliocentric energy and asymptotic direction, after which the spacecraft coasts ballistically and arrives with a residual velocity mismatch that must be removed during capture. This requires planetocentric $\Delta v$ not only to eliminate the chosen $v_{\infty,\text{planet}}$ but also to re-orient and reduce the velocity vector to achieve the desired science orbit.

However, the steep decay of solar power with heliocentric distance makes SEP ill-suited to short, high-$\Delta v$ captures at Uranus' orbit, so a fast SEP--SOM transfer must either constrain $v_\infty$ to levels manageable by a moderate chemical stage or employ additional technologies such as aerocapture \cite{Pradeepkumar2023fast}. The large deployable array might additionally serve as a drag surface for aerocapture or aerobraking at atmosphere-bearing planets, though its thermal and structural survivability in that role remains to be assessed. Alternatively, if nuclear systems are permitted, the same EP thrusters could be powered by a smaller nuclear power source for a prolonged low-thrust spiral capture after the initial SEP--SOM. A full trade of these design options is deferred to future work.

\subsection{Validity bounds and limitations}\label{subsec:limitations}

The results are conditional on the simplified feasibility model summarised in Section~\ref{subsec:scope_limits} and on the reference technology point used to generate the baseline trajectories. The dominant dependency is the assumed system-level EPS specific power at elevated temperature. The reference case with $\alpha_{\mathrm{EPS}}=200~\mathrm{W\,kg^{-1}}$ is an illustrative anchor rather than a universal feasibility requirement. The sensitivity maps in Section~\ref{subsec:sensitivity_analysis} quantify how payload degrades with lower $\alpha_{\mathrm{EPS}}$ and higher $\mu_s$ and provide the corresponding threshold interpretation across lower and intermediate specific-power levels. The dynamics and operations are idealised, using planar two-body propagation, an analytical in-plane JGA when applicable, and continuous thrust availability. The high-power EP system is abstracted by constant $I_{sp}$ and $\eta$ enabled by thruster-string switching over the power envelope. Finally, SOMBRERO is a stochastic optimiser, so the reported steering laws should be interpreted as near-optimal solutions for the chosen architecture and control parametrisation, not strict upper bounds.

%=======================================================================

\section{Conclusions and Future Work}
\label{sec:conclusion}

We have presented an architecture-level feasibility study of a high-payload Solar Electric Propulsion Solar Oberth Maneuver (SEP--SOM) for deep-space escape, designed to reduce reliance on nuclear-powered propulsion and to address key limitations of existing deep-space escape concepts such as limited payload capability, complex multi-assist gravity-assist sequences, or the need for costly super-heavy launchers.
This design exploits the Oberth effect, using emerging high-temperature solar cells to power high-power electric propulsion at close solar distances.

Our analysis focuses on a SEP--SOM mission that concentrates low-thrust burns at $\sim0.3~\mathrm{AU}$ to maximise payload delivery to $200~\mathrm{AU}$ within $25~\mathrm{yr}$. Based on laboratory demonstrations of high-temperature solar cells, we adopt a thermal survivability assumption of $\sim400\,^{\circ}\mathrm{C}$ and use a reference EPS specific power of $\alpha_{\mathrm{EPS}}=200~\mathrm{W\,kg^{-1}}$ at $1~\mathrm{AU}$ as an illustrative high-performance baseline motivated by published development targets for conventional arrays. Spacecraft configuration and steering were optimised using the SOMBRERO evolutionary algorithm under a planar model, and a sensitivity analysis was then used to map payload-feasibility thresholds across lower $\alpha_{\mathrm{EPS}}$ values.

Key findings from the baseline analysis are as follows. First, an optimised SEP--SOM (expendable Falcon Heavy launch) can deliver $\approx3{,}080~\mathrm{kg}$ to $200~\mathrm{AU}$ in $\approx25.0~\mathrm{yr}$ (with a Jupiter gravity assist) or $\approx1{,}550~\mathrm{kg}$ in $24.3~\mathrm{yr}$ direct. This allocatable payload is nearly two orders of magnitude above the 35~kg instrument mass of the closest comparable prior solar-electric Oberth concept (Table~\ref{tab:som_literature}). Second, concentrating thrust near the Sun yields roughly a $3\times$ increase in delivered specific orbital energy versus a $1~\mathrm{AU}$ spiral for the same $\Delta v$. This is the primary mechanism behind the SEP--SOM performance gain. Third, the sensitivity maps show that, for the illustrated structural mass fraction case ($\mu_s=0.30$), payload-feasibility thresholds sit near present-day conventional EPS levels for JGA cases and above them for direct cases. Relative to the present-day order-of-magnitude EPS-level value ($\sim78.8~\mathrm{W\,kg^{-1}}$), a $100~\mathrm{kg}$ payload is feasible at roughly $0.9\times/1.5\times$ (JGA/direct), while ton-class payloads ($\geq 1{,}000~\mathrm{kg}$) become feasible at roughly $1.1\times/2.0\times$ (JGA/direct). These results define an architecture-level feasibility envelope for high-payload heliopause missions without nuclear propulsion or costly super-heavy launchers (e.g. SLS) under the stated model assumptions.

At the architecture level, these results suggest that HIHT solar cells could shift from survival hardware to propulsion-enabling technology for near-Sun electric propulsion, providing a novel technology-maturation rationale. Using SEP only as a short, high-power phase (front-loading $\Delta v$) offers a potential non-nuclear (electric) propulsion approach for interstellar precursors. This near-Sun energy front-loading concept may also apply to fast outer-planet or high-$\Delta v$ missions. Under the reference-case assumptions, a single near-Sun SEP--SOM could thus provide a high-payload precursor trajectory to the heliopause with one \emph{commercial} launcher (here, an expendable Falcon Heavy).

Future work should focus on three extensions. First, translate EPS assumptions from cell-level HIHT demonstrations to panel- and system-level designs, including temperature-dependent degradation, integration penalties, and explicit margins. Second, extend the optimisation to multi-architecture trades across payload, transfer time, perihelion distance, launch energy, and alternative high-power EP options to identify regimes where SEP--SOM is competitive compared with multi-assist or nuclear electric solutions. Third, move from planar two-body dynamics to 3D, ephemeris-based optimisation with operational and navigation constraints to quantify robustness to dispersions and feasibility under guidance, thermal, and pointing limits.

Together, these steps would help translate the trajectory-level performance trends established here into more detailed mission designs. If the assumed HIHT array performance can be realised at panel level with credible margins, we suggest that low-perihelion SEP--SOM be considered among candidate architectures for interstellar-precursor and other high-energy deep-space missions.


%% back matter: let columns end short rather than stretching glue to fill them
%% (the abbreviations/symbols tables are rigid boxes, so \flushbottom opens gaps)
\raggedbottom
\section*{Abbreviations}
\label{sec:abbreviations}

\noindent
\begin{tabular}{@{} p{1.2cm} p{5.9cm} @{}}AF-MPD & Applied-Field Magnetoplasmadynamic Thruster \\
ANN    & Artificial Neural Network \\
COTS   & Commercial Off-The-Shelf \\
EGA & Earth Gravity Assist\\
ENC    & Evolutionary Neuro-Control \\
EP     & Electric Propulsion \\
EPS    & Electrical Power System \\
GIT    & Gridded Ion Thrusters \\
HIHT   & High-Intensity High-Temperature \\
JGA    & Jupiter Gravity Assist \\
LEO    & Low Earth Orbit \\
NTP    & Nuclear Thermal Propulsion \\
NTR    & Nuclear Thermal Rocket \\
PPU    & Power Processing Unit \\
REP    & Radioisotope Electric Propulsion \\
RTG    & Radioisotope Thermoelectric Generator \\
SEP    & Solar Electric Propulsion \\
SOM    & Solar Oberth Maneuver \\
SLS    & Space Launch System \\
STP    & Solar Thermal Propulsion \\
TRL    & Technology Readiness Level \\
VGA    & Venus Gravity Assist \\
\end{tabular}

\section*{Symbols}
\label{sec:symbols}

% Header setup for Symbols table
\noindent
\begin{tabular}{@{} p{1.2cm} p{4.8cm} p{1.7cm} @{}}
\textbf{Sym.} & \textbf{Definition} & \textbf{Unit} \\
$a$                & semi-major axis                & [AU]  \\
$\boldsymbol{c_0}$ & initial conditions             & [--]  \\
$C_3$              & characteristic energy          & [km$^2$\,s$^{-2}$]  \\
$F_T$              & thrust magnitude               & [N]  \\
$GM$               & gravitational parameter        & [m$^3$\,s$^{-2}$]  \\
$I_0$              & solar irradiance at 1~AU       & [W\,m$^{-2}$]  \\
$I_{sp}$           & specific impulse               & [s]  \\
$J$                & fitness score                  & [--]  \\
$j_i$              & sub-fitness scores             & [--]  \\
$m_0$              & launch mass                    & [kg]  \\
$m_{dry}$          & delivered dry spacecraft mass  & [kg]  \\
$m_{EPS}$          & mass of EPS                    & [kg]  \\
$m_p$              & propellant mass                & [kg]  \\
$m_{pl}$           & allocatable mission payload mass & [kg]  \\
$m_s$              & structure mass                 & [kg]  \\
$m_{sc}$           & instantaneous spacecraft mass  & [kg]  \\
$P$                & power                          & [W]  \\
$P(\chi)$          & steering policy                & [--]  \\
$p_m$              & mutation probability           & [--]  \\
$R_\odot$          & solar radius                   & [km]  \\
$r_{JGA}$          & periapsis distance for JGA     & [km]  \\
$r_{SOM}$          & distance at SOM perihelion     & [AU]  \\
$T$                & temperature                    & [K]  \\
$t_{\text{200AU}}$ & flight time to 200 AU          & [years]  \\
$\boldsymbol{u}$   & steering control commands      & [--]  \\
$u_P$              & throttle command               & [--]  \\
$v$                & orbital velocity               & [km\,s$^{-1}$]  \\
$v_{escape,J}$     & solar escape speed at 5.2 AU   & [m\,s$^{-1}$]  \\
$v_\infty$         & hyperbolic excess velocity     & [AU\,year$^{-1}$]  \\
\end{tabular}

\medskip

\noindent
\begin{tabular}{@{} p{1.2cm} p{4.8cm} p{1.7cm} @{}}
\textbf{Sym.} & \textbf{Definition} & \textbf{Unit} \\
$\alpha$           & specific power                 & [W\,kg$^{-1}$]  \\
$\gamma$           & decay parameter                & [s$^2$\,km$^{-2}$]  \\
$\Delta v$         & orbital velocity increment     & [km\,s$^{-1}$]  \\
$\delta$           & turn angle for gravity assist  & [rad]  \\
$\epsilon$         & emissivity                     & [--]  \\
$\varepsilon$      & specific orbital energy        & [J\,kg$^{-1}$]  \\
$\eta$             & efficiency                     & [--]  \\
$\theta_T$         & thrust angle                   & [rad]  \\
$\kappa$           & scaling exponent for EPS power & [--]  \\
$\mu$              & mass fraction                  & [--]  \\
$\nu$              & true anomaly                   & [rad]  \\
$\boldsymbol{\xi}$ & ANN genetic material parameter & [--]  \\
$\boldsymbol{\pi}$ & ANN genetic chromosome         & [--]  \\
$\sigma$           & activation functions           & [--]  \\
$\varphi$          & angle between velocity and position vector              & [rad]  \\
$\chi$             & spacecraft state               & [--]  \\
\end{tabular}

\section*{Author contributions}
\begin{itemize}
 \item \textbf{N.M.:} Conceptualisation; Project Administration; Methodology; Software; Formal analysis; Investigation; Data curation; Visualisation; Writing -- original draft; Writing -- review \& editing.
  \item \textbf{W.v.L.:} Investigation (literature review on high-temperature solar arrays and electric propulsion technologies); Writing -- original draft (background subsections, discussion subsections); Writing -- review \& editing.
 \item \textbf{C.G.O.B.:} Investigation (literature review on Solar Oberth maneuvers and high-temperature solar cells); Visualisation; Writing -- original draft (background subsections, discussion subsections); Writing -- review \& editing.
 \item \textbf{A.H.:} Writing -- review \& editing.
\end{itemize}

\section*{Data availability}
The curated supporting artifacts (Excel trajectory time-series and scalar summaries 
for the JGA and no-JGA cases) are available in the public
SOMBRERO repository at \url{https://github.com/astronadim/sombrero-trajectory-optimizer} under \texttt{data/paper\_artifacts/}.
Primary reproduction of the reported reference 
trajectories is via the configuration files and optimised chromosomes provided 
under \texttt{configs/}, using the \texttt{simulate} command documented in the 
repository README.

\section*{Code availability}
The SOMBRERO trajectory optimisation and simulation code used in this study is
publicly available under the MIT License at \url{https://github.com/astronadim/sombrero-trajectory-optimizer} (release v1.0.0). The repository
includes configuration files, optimised neural-network chromosomes, and 
installation instructions for hot-start reconstruction of the two paper reference 
trajectories (Tables~\ref{tab:SOM_JGA_results_summary} and~\ref{tab:results_summary_no_JGA}). Reproducibility scope and platform-dependence 
notes are provided in the repository documentation.
\section*{Competing interests}
The authors declare no competing interests.

\section*{Acknowledgements}
This research did not receive any specific grant from funding agencies in the public, commercial, or not-for-profit sectors.

\appendix

\phantomsection
\pdfbookmark[0]{Appendices}{app:start}

\section{Supplementary literature review: non-electric SOM propulsion}
\label{app:non_ep_som_propulsion}
This appendix provides additional discussion of non-electric SOM implementations (chemical, solar sail, solar thermal, and nuclear thermal), complementing Table~\ref{tab:som_literature}. The main text focuses on electric-propulsion SOM architectures.

\subsection{Chemical Propulsion SOM}

Chemical Solar Oberth Maneuvers have been studied for high-energy, long-range missions (see Table~\ref{tab:som_literature}), with targets ranging from the heliopause ($\sim$100~AU) \cite{hopkins2015propulsion} to encounters with the hypothetical Planet~9 ($\sim$400~AU) \cite{hibberd2022can} or interstellar objects such as 3I/Atlas \cite{hibberd2026catching3iatlasusingsolar}. Across these scenarios, mass constraints remain the principal limitation. Reported delivered payloads vary from $\sim$100~kg \cite{hibberd2022can} to $\sim$500~kg \cite{hibberd2026catching3iatlasusingsolar}, though in reality these numbers would be reduced by the parasitic heat shield mass. 
For instance, in Hopkins et al.'s concept~\cite{hopkins2015propulsion}, the reported delivered dry spacecraft mass is $\sim$380~kg, while the low perihelion (11 solar radii) thermal environment for the SOM requires a $\sim$300~kg heat shield. In comparison, perihelion conditions in the scenarios reported by Hibberd et al.~\cite{hibberd2022can} and Hibberd \& Eubanks \cite{hibberd2026catching3iatlasusingsolar} are even more extreme, implying an even larger fraction of the payload could be consumed by thermal protection.

\subsection{Solar Sail Propulsion SOM}

Highly reflective solar sails can harness radiation pressure to accelerate spacecraft, with closer perihelia providing higher $\Delta v$ through a Solar Oberth Maneuver \cite{bailer2021sun}. Lightweight sails with large area-to-mass ratios could enable velocities exceeding $20~\mathrm{AU\,year^{-1}}$, more than 5 times the cruise speed of Voyager~1, for perihelia as low as 0.05~AU. Mass remains a key constraint, often requiring sail areas $>10{,}000$~m$^2$ even for delivered dry spacecraft masses of 10--50~kg \cite{davoyan2021photonic, bailer2021sun}. Concepts such as a $\sim$400~m sail carrying a 25~kg \emph{science payload} illustrate the scale of the required deployable structures (see Table~\ref{tab:som_literature}).

Solar sailing has progressed from concept to demonstration, including IKAROS (310~kg spacecraft, 200~m$^2$ sail), NanoSail-D, LightSail~1/2, and the unflown NASA Solar Cruiser ($\sim$100~kg, 1,700~m$^2$) \cite{bailer2021sun,johnson2023solarsailpropulsion2050}. Future proposals suggest extreme solar flybys (<5 \,$R_\odot$) could accelerate 10--20~kg spacecraft to $\sim 60~\mathrm{AU\,year^{-1}}$ ($\approx 0.001c$), reaching Neptune in 10 months or 1{,}000~AU in 17 years, with swarm formation enabling larger effective payloads \cite{davoyan2024extreme}.

Laser-driven sails offer an alternative, using Earth-based high-power lasers for acceleration. Hibberd et al. \cite{hibberd2020projectlyracatching1ioumuamua} studied gram- to kilogram-scale spacecraft concepts for intercepting 1I/‘Oumuamua using laser-accelerated sailcraft at velocities of 0.001 c. Their analysis indicates a minimum flight duration of approximately 444 days, reaching interception distances beyond 82 AU, assuming a perihelion constraint of 3 solar radii. Payload capacity and required infrastructure remain key limiting factors.

\subsection{Solar Thermal Propulsion (STP)}

Solar Thermal Propulsion (STP) converts radiant solar energy into thermal energy, which is then transformed into kinetic energy to provide thrust, offering at least double the specific impulse of chemical propulsion under conservative assumptions~\cite{benkoski2023final} and enabling high-energy Solar Oberth Maneuvers at low perihelia \cite{lyman2001_stp_interstellar, shoji1992_solarthermal} (see Table~\ref{tab:som_literature}).

Early conceptual studies demonstrated velocities above $11~\mathrm{AU\,year^{-1}}$ for perihelia $\sim$0.02~AU~\cite{shoji1992_solarthermal}, while later designs, such as a 50~kg probe performing a Jupiter flyby and perihelion burn at 3--4~$R_\odot$, could reach $20~\mathrm{AU\,year^{-1}}$~\cite{lyman2001_stp_interstellar}. Sauder et al.~\cite{sauder2021system} showed that a 478~kg probe could achieve $8-10~\mathrm{AU\,year^{-1}}$ with a 36~kg \emph{science payload}, and hybrid architectures combining STP with nuclear-electric or advanced ion propulsion could push velocities to $19.5-21~\mathrm{AU\,year^{-1}}$. Benkoski et al.~\cite{benkoski2023combined} experimentally demonstrated hardware potentially capable of $15~\mathrm{AU\,year^{-1}}$ for 2.5~$R_\odot$ perihelia, though payload remains heavily constrained by thermal protection and cooling requirements.

\subsection{Nuclear Thermal Propulsion SOM}

Nuclear Thermal Propulsion (NTP) has been studied for fast interstellar precursor missions, combining SLS launches with perihelion Solar Oberth Maneuvers (see Table~\ref{tab:som_literature}). Hibberd \& Hein~\cite{hibberd2021project} analyse fast intercept missions to interstellar object 1I/‘Oumuamua, with a potential LEO $\rightarrow$ SOM $\rightarrow$ 1I trajectory achieving 1.3~t (no JGA) to 7.8~t (with JGA) spacecraft masses over 21 to 32 years, though residual mass available for payload is very limited. D. Scott~\cite{scott2018analysis} evaluates NTR missions to the heliopause and beyond, showing that an SLS Block 1 plus a 10~$R_\odot$ perihelion burn can reach $43-59.8~\mathrm{km\,s^{-1}}$ exit velocities; at constant speed these give 500~AU in roughly 40--55~years, again with limited payloads. Edwards et al. \cite{irvine2020design} examine various combinations of launch vehicles, propellant choices, and flyby sequences. They identify a maximum escape velocity scenario, achieving an exit velocity of 14.8 AU.yr$^{-1}$ and a flight time to 100 AU of 16.4 years. In contrast, they also present a minimum flight time scenario of 10.1 years, albeit with a reduced exit velocity of 13.9 AU.yr$^{-1}$.

\section{Supplementary material: SOMBRERO implementation and supporting models}
\label{app:sombrero}
This appendix collects implementation-level details needed to reproduce the SOMBRERO results. Neurocontrol and fitness definitions are given in \ref{app:enc_details} and \ref{app:fitness}. Supporting models and fits are documented in \ref{app:launch}, \ref{app:messenger_mus}, \ref{app:rsom_scaling}, and \ref{app:jga_model}. Verification and validation are reported in \ref{app:validation_InTrance}; the no-JGA ablation case is covered in \ref{app:nojga_case}.

\subsection{Evolutionary neurocontrol method definitions}
\label{app:enc_details}

\subsubsection{Activation functions}

The Gaussian-type activation functions 
\begin{equation}
\sigma_{\theta_{T}}(x)=-\exp(-x^2),\quad \sigma_{u_{P}}(x)=1-\exp(-5x^2)
\end{equation}
in the output layer, along with the $\sigma_h(x)=\tanh(x)$ functions in the hidden layer, were chosen to aid learning.
$\sigma_{\theta_{T}}$ and $\sigma_{u_{P}}$ are designed to match the expected behaviour of $u_{P} \rightarrow 100\%$ and $\theta_{T} \rightarrow 0$ in the most critical point of the trajectory, the SOM perihelion, when receiving similar input states. This simplifies learning by limiting the solution space to negative $\theta_T$. In the implementation, the raw network throttle output $\sigma_{u_{P}}(x)\in[0,1)$ is linearly rescaled by a factor $1/0.8$ and saturated to give the realised command $u_{P}=\min(\sigma_{u_{P}}/0.8,\,1)\in[0,1]$, so raw outputs above $0.8$ command full power.

\subsubsection{Crossover}
Crossover is a genetic operator used to combine the genetic information of two parents to produce offspring. In the SEP--SOM context, each offspring is produced by one of the four operators below, drawn with equal probability (25\%) \cite{dachwald2004evolutionary}:
\begin{itemize}
\item \textbf{One Point Crossover:} A single crossover point on each chromosome is chosen. The data beyond that point in either chromosome is swapped between the two parent chromosomes.
\item \textbf{Uniform Crossover:} Each gene in the offspring is randomly chosen from the corresponding genes of the parents, with equal probability.
\item \textbf{Node Crossover:} Nodes from both parents are randomly selected and swapped, helping the offspring inherit structural properties from both parents.
\item \textbf{Arithmetic Crossover:} Offspring are created by performing a weighted average of the parent chromosomes. The weight is randomly selected.
\end{itemize}
Following best practice, information from the winner is retained: only nodes and node information (weights, biases) are exchanged \cite{ohndorf2016multiphase}.

\subsubsection{Jupiter Gravity Assist Model}\label{app:jga_model}

A Jupiter Gravity Assist (JGA) maneuver is implemented as described by \cite{walter2008astronautics}. The process is summarised below:

In this planar analysis, the JGA is modelled as an instantaneous change in the spacecraft's heliocentric velocity at Jupiter's orbital radius. This simplification neglects three-dimensional effects and finite-duration flyby arcs but captures the principal energy gain mechanism. The periapsis distance of the JGA is fixed, and the spacecraft's approach is assumed to occur at Jupiter's orbital radius. The JGA periapsis is held fixed at 1.34 Jupiter radii and is not optimised, which keeps the assist consistent across cases. The reported energy gain is representative of a favourably phased encounter.

The spacecraft's position and velocity are represented by $(x, y)$ and $(\dot{x}, \dot{y})$, respectively, in a heliocentric inertial frame. The relative velocity of the spacecraft with respect to Jupiter is computed as:
\begin{equation}
  \mathbf{v_{\text{relative}}} = \mathbf{v_{\text{sc}}} - \mathbf{v_{\text{Jupiter}}}
\end{equation}

where $\mathbf{v_{\text{sc}}}$ is the spacecraft's heliocentric velocity and $\mathbf{v_{\text{Jupiter}}}$ is Jupiter's heliocentric velocity vector. The turn angle $\delta$ for the gravity assist is determined by:
\begin{equation}\label{eq:JGA_rotation_angle}
  \delta = 2 \arcsin\left(\frac{1}{1 + \frac{r_{\text{JGA}} \cdot \|v_{\text{relative}}\|^2}{GM_{\text{Jupiter}}}}\right)
\end{equation}

where $r_{\text{JGA}}$ is the periapsis distance of the JGA, $v_{\text{relative}}$ is the magnitude of the spacecraft's velocity relative to Jupiter, and $GM_{\text{Jupiter}}$ is the gravitational parameter of Jupiter.

A rotation matrix is applied to the relative velocity vector to compute the final velocity after the JGA:

\begin{equation}
  \mathbf{R_\delta} = 
  \begin{bmatrix}
  \cos(\delta) & -\sin(\delta) \\
  \sin(\delta) & \cos(\delta)
  \end{bmatrix}
\end{equation}

\begin{equation}
  \mathbf{v_{\text{final}}} = \mathbf{R_\delta} \cdot \mathbf{v_{\text{relative}}} + \mathbf{v_{\text{Jupiter}}}
\end{equation}

where $\mathbf{v_{\text{final}}}$ is the final velocity of the spacecraft after the gravity assist, and $\mathbf{v_{\text{Jupiter}}}$ is Jupiter's heliocentric velocity vector. The periapsis distance $r_{\text{JGA}}$ is assumed to be 1.34 Jupiter radii \cite{ohndorf2016multiphase}.

\subsection{MESSENGER spacecraft structural mass fraction estimate}
\label{app:messenger_mus}
SMAD Table~A-10 reports for MESSENGER $m_0=1102~\mathrm{kg}$, $m_{\mathrm{dry}}=508~\mathrm{kg}$, $m_{\mathrm{pl}}=47~\mathrm{kg}$, and an electric power system mass share of $\tilde{\mu}_{\mathrm{EPS}}=0.24$ given as a fraction of dry mass \cite{WertzEverettPuschell2011NewSMAD}.
Using the paper bookkeeping $m_{\mathrm{dry}}=m_{\mathrm{pl}}+m_{\mathrm{EPS}}+m_s$, we set $m_{\mathrm{EPS}}=\tilde{\mu}_{\mathrm{EPS}}\,m_{\mathrm{dry}}$ and obtain the launch-mass structural fraction
\begin{equation}
\begin{aligned}
\mu_{s}
&\equiv\frac{m_s}{m_0}
=\frac{m_{\mathrm{dry}}-m_{\mathrm{pl}}-\tilde{\mu}_{\mathrm{EPS}}\,m_{\mathrm{dry}}}{m_0} \\
&=\frac{508-47-0.24\cdot 508}{1102}=0.30770.
\end{aligned}
\end{equation}

%%%%%%%%%%%%%

\subsection{Temperature-limited perihelion scaling}
\label{app:rsom_scaling}
This appendix summarises the radiative-equilibrium argument used to relate a temperature limit of the photovoltaic system to a minimum feasible SEP--SOM perihelion distance.

We model the incident solar irradiance as $I(r)=I_0/r^2$, with $r$ in astronomical units and $I_0$ the solar constant at 1~AU. For a panel with net absorbed heat flux $\alpha_{\mathrm{net}} I(r)$, radiative equilibrium with emission from the front and rear sides gives \cite{landis2008solar,genovese2023advanced}
\begin{equation}
\alpha_{\mathrm{net}}\,\frac{I_0}{r^2} = (\epsilon_f+\epsilon_r)\,\sigma\,T^4,
\label{eq:rsom_balance}
\end{equation}
where $\epsilon_f$ and $\epsilon_r$ are the front- and rear-side hemispherical emissivities, $\sigma=5.670374419\times10^{-8}\,\mathrm{W\,m^{-2}\,K^{-4}}$ is the Stefan--Boltzmann constant, $T$ is the panel operating temperature (in K), and $\alpha_{\mathrm{net}}$ denotes the net fraction of incident irradiance converted to heat (absorptance minus electrical conversion efficiency). Rearranging yields the scaling
\begin{equation}
r_{\mathrm{SOM}} = \sqrt{\frac{\alpha_{\mathrm{net}}}{\epsilon_f+\epsilon_r}\,\frac{I_0}{\sigma}}\;\frac{1}{T^2},
\label{eq:rsom_scaling}
\end{equation}
i.e.\ $r_{\mathrm{SOM}}\propto T^{-2}$ for fixed optical properties.

A commonly used simplifying case is $\alpha_{\mathrm{net}}=\epsilon_f=\epsilon_r$, corresponding to symmetric emission from both sides. In contrast, if only one side effectively emits, the denominator reduces from $(\epsilon_f+\epsilon_r)$ to a single emissivity, which increases the implied perihelion distance by a factor $\sqrt{2}$ for the same $T$ and $\alpha_{\mathrm{net}}$. In the main text, this range is used only to motivate the representative baseline choice $r_{\mathrm{SOM}}=0.3\,\mathrm{AU}$ for $T\approx400^\circ\mathrm{C}$.
 
%%%%%%%%%%%%%

%%%%%%%%%%%%%
\subsection{Fitness function formulations and heuristics}
\label{app:fitness}

This appendix provides the detailed mathematical definitions and heuristic filters used within the evolutionary neurocontrol framework, as introduced in Section~\ref{subsec:neurocontrol}.

\subsubsection{Heuristic Filter for SOM Candidates}
To accelerate convergence and prune the search space of unfit individuals, a heuristic filter is applied after the first generation. Only individuals showing viable SOM characteristics (termed ``SOM candidates'') are allowed to ``survive'' and be passed to the next generation. Individuals that fail to meet these criteria are re-initialised. This pre-selection process, based on characteristics expected from a successful SOM trajectory, reduces dependence on the randomly initialised first generation.

The criteria for this filter are:
\begin{itemize}
    \item[(a)] Average heliocentric distance $r_{avg} > 1.2~\mathrm{AU}$.
    \item[(b)] The trajectory must contain only one local maximum in $r$.
    \item[(c)] The local maximum in $r$ must be greater than 2 AU.
    \item[(d)] The final radial distance $r_{final} > 1$ AU after 4.5 years of simulation.
\end{itemize}

\subsubsection{Staged Fitness Function Definitions}
SOM candidates enter the main evolutionary loop, where their performance is evaluated based on the five successive stages described in the main text. The mathematical formulation for each sub-fitness score $j_i$ is as follows:

\begin{enumerate}
    \item \textbf{Solar Oberth Maneuver:} This phase evaluates the spacecraft's ability to achieve a minimal distance $r_{min}$ that matches the design perihelion $r_{SOM,design}$, with penalties for deviations. This evolutionary stage is considered passed if $0.25~\mathrm{AU} < r_{min} < 0.7~\mathrm{AU}$.
    \begin{equation} \label{eq:j1}
        j_{1}(r_{min}) = 
        \begin{cases}
            \exp(-1000(r_{min} - r_{SOM,design})^{2}) & \text{if } r_{min} < r_{SOM,design} \\
            \exp(-50(r_{min} - r_{SOM,design})^{2}) & \text{if } r_{min} \ge r_{SOM,design}
        \end{cases}
    \end{equation}

    \item \textbf{Approach Jupiter:} After the SOM, the $j_2$ score rewards reaching a maximal radial distance $r_{max}$ near Jupiter's orbit (5.2 AU).
    \begin{equation} \label{eq:j2}
        j_{2}(r_{max}) = 
        \begin{cases}
            \frac{r_{max}}{5.2} & \text{if } r_{max} < 5.2~\mathrm{AU} \\
            1 & \text{if } r_{max} \ge 5.2~\mathrm{AU}
        \end{cases}
    \end{equation}

    \item \textbf{Reaching Solar Escape Velocity:} After the JGA, the spacecraft velocity $v$ must be at or above the escape velocity from Jupiter's orbit (5.2~AU) $v_{\text{escape,J}} = 18471~\mathrm{m\,s^{-1}}$. The $j_3$ score rewards kinetic energies approaching this threshold.
    \begin{equation} \label{eq:j3}
        j_{3}(v) = 
        \begin{cases}
            \frac{v^{2}}{v_{\text{escape,J}}^{2}} & \text{if } v < 18471~\mathrm{m\,s^{-1}} \\
            1 & \text{if } v \ge 18471~\mathrm{m\,s^{-1}}
        \end{cases}
    \end{equation}

    \item \textbf{Time to 200 AU:} The spacecraft must reach 200 AU within 25 years. The $j_4$ score rewards smaller $t_{200AU}$. This stage is considered passed when 200 AU is reached within the 25-year target time.
    \begin{equation} \label{eq:j4}
        j_{4}(t_{200AU}) = 
        \begin{cases}
            -\tanh(0.5(\frac{t_{200AU}}{25~\text{years}} - 1)) + 1 & \text{if } t_{200AU} > 25~\text{years} \\
            1 & \text{if } t_{200AU} \le 25~\text{years}
        \end{cases}
    \end{equation}

    \item \textbf{Maximise Payload Mass:}  This final phase, activated once stages 1--4 have been passed ($j_2=j_3=j_4=1$), focuses on maximising the payload mass $m_{pl}$. The score $j_5$ is equivalent to $m_{pl}$ in units of 1000 kg.
    \begin{equation} \label{eq:j5}
        j_{5}(m_{pl}) = m_{pl} / 1000~kg
    \end{equation}
\end{enumerate}

%%%%%%%%%%%

\subsection{Launch}\label{app:launch}

\begin{figure}[h!]
    \centering
    \includegraphics[width=\linewidth]{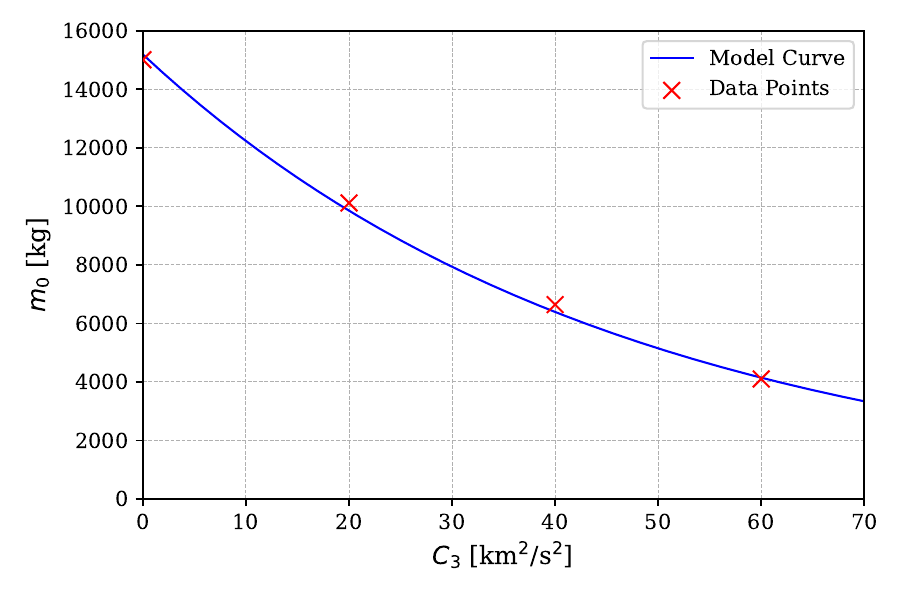}
    \caption{\textbf{Injected mass model for an expendable Falcon Heavy as a function of launch energy $C_3$, used in this study.}}

    \label{fig:app_c3_payload}
\end{figure}

Using Falcon Heavy (expendable) data from \cite{nasa_elv_performance} a fit of 
\begin{equation}
    m_0=m_{0,max} \exp(-\gamma \cdot C_3)
\end{equation} 

yields $m_{0,max} = 15{,}189$~kg and $\gamma=0.02165~\mathrm{s^2\,km^{-2}}$, as shown in Fig.~\ref{fig:app_c3_payload}.

%%%%%%%%%%%%%%%
\subsection{Validation}\label{app:validation_InTrance}

\subsubsection{Rebuilding Literature Results}\label{app:trajectory_validation}

Initial conditions from \cite{ohndorf2011flight,loeb2011interstellar} are shown in Table~\ref{tab:initial_conditions}.

\begin{table}[ht]
\centering
\caption{\textbf{Initial and boundary conditions for the literature reference SOM case with $r_{\mathrm{SOM}}=0.7~\mathrm{AU}$ used in the validation study.}}
\label{tab:initial_conditions}
\begin{tabular}{lclc}
\hline
\textbf{Parameter} & \textbf{Value} & \textbf{Parameter} & \textbf{Value} \\
\hline
$I_{sp}$ & 7,377 s & Launcher & Ariane 5 \\
$\eta$ & 71.6\% & $C_3$ & $45.1~\mathrm{km^2\,s^{-2}}$ \\
$P_0$ (1 AU) & 53 kW & $\mu_{\text{EPS}}$ & 0.157 \\
$\alpha$ & $200~\mathrm{W\,kg^{-1}}$ & $\mu_s$ & 0.289 \\
\hline
\end{tabular}
\end{table}

The subsystem mass fraction, $\mu_s$, stems from the assumed fixed-thrust units (6 RIT-22 + power units), the Xenon tank, and the supporting structure, which together amount to
\begin{equation}
  m_s = \SI{224}{kg} + \SI{26}{kg} + \SI{239}{kg} = \SI{489}{kg}
\end{equation}
leading to $\mu_s=0.289$ (489~kg out of 1{,}692~kg) \cite{loeb2011interstellar}. Note that Loeb et al.\ do not treat the combined propellant and payload mass fraction at launch,
\begin{equation}
  \mu_{pp,0} = 1 - \mu_s - \mu_{EPS}
\end{equation}
as part of the optimised initial conditions $\mathbf{c}_0$. Instead, they assume a characteristic power fraction of 
\begin{equation}
  P_c = 65\%,
\end{equation}
which is defined as the percentage of power at 1~AU relative to the maximum power available at perihelion, in the case of the six RIT22 thrusters: $P_{\max}=6\cdot 13.59\,\text{kW}$. This assumption determines the electrical power system (EPS) mass fraction as
\begin{equation}
\mu_{\text{EPS}} = \frac{P_{\max} \, P_c}{\alpha \, m_0}
= \frac{81.54\,\text{kW}\cdot 0.65}{200~\mathrm{W\,kg^{-1}}\cdot 1{,}692\,\text{kg}}
\approx 0.157.
\end{equation}
The underlying $\alpha = 200~\mathrm{W\,kg^{-1}}$ is the array-level specific power ($5~\mathrm{kg\,kW^{-1}}$) adopted by Loeb et al.\ for the solar array \cite{loeb2011interstellar}; it coincides numerically with, but is distinct in integration level from, the EPS-level reference anchor $\alpha_{\mathrm{EPS}}$ of this study. In the reproduced design, the power-processing units are contained in the $224$~kg thrust-unit mass and thus in $m_s$.

\begin{figure}[h!]
    \centering
    \includegraphics[width=\linewidth]{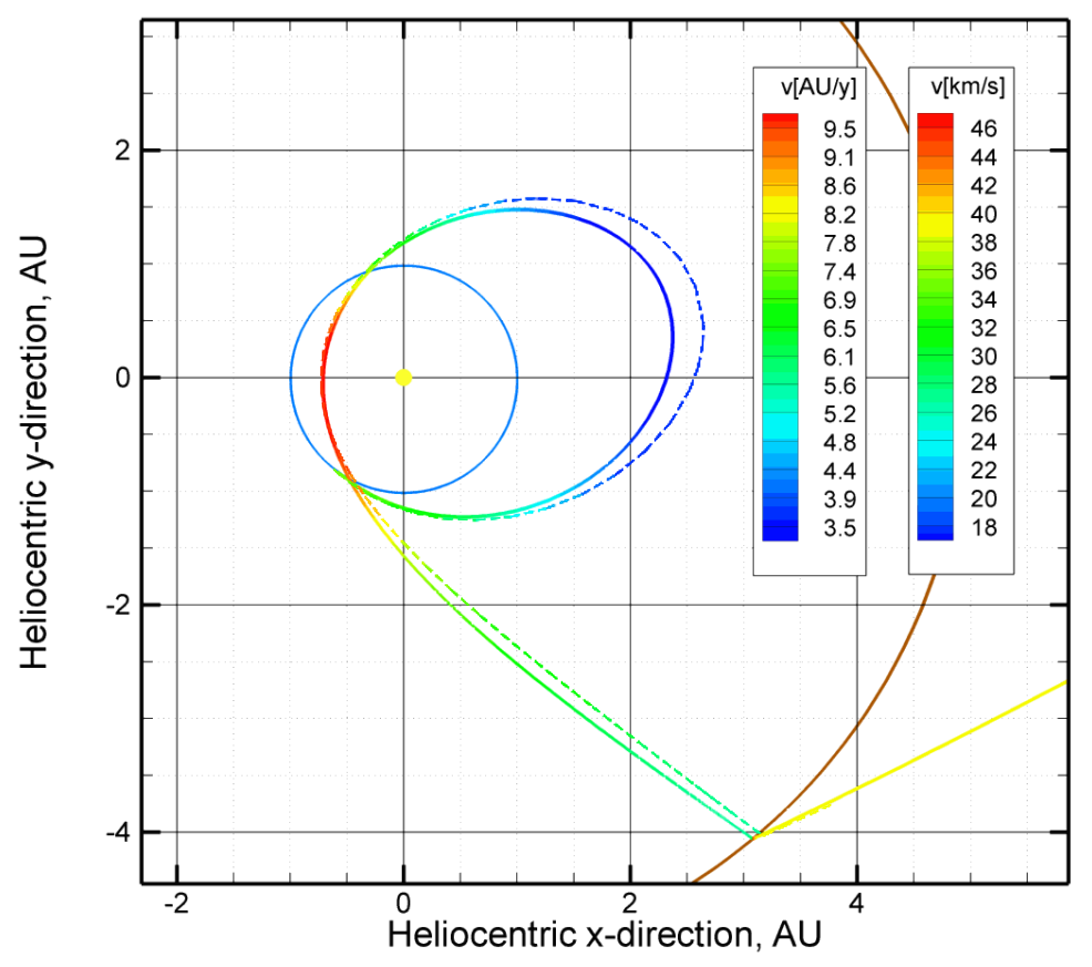}
    \caption{\textbf{Overlay of heliocentric in-plane trajectories for the reference SOM case.} Solid line: InTrance result from Ohndorf et al. \cite{ohndorf2011flight}. Dashed line: SOMBRERO reproduction under the same initial conditions.}
    \label{fig:comparison}
\end{figure}

Trajectory from \cite{ohndorf2011flight}, overlaid with the trajectory from the SOMBRERO tool, is shown in Fig.~\ref{fig:comparison}. The optimised trajectory (dashed line) here is close to the one reported by Ohndorf et al. (solid line). Small deviations in the aphelion result in differences in the final payload and propellant masses (see Table~\ref{tab:comparison_validation_results}). These deviations likely come from differences in the optimisation approaches. InTrance uses a high-fidelity model that simulates Earth and Jupiter as moving bodies, so the trajectory must actively "catch" Jupiter for a gravity assist. SOMBRERO considers only the orbital radii and applies an analytical gravity assist once the spacecraft reaches Jupiter's orbit at 5.2 AU.

\subsubsection{Thrust Acceleration Error Assessment}\label{app:thrust_error}

A further confirmation of the simulation accuracy is provided by a detailed thrust acceleration error assessment. The analytical velocity increment is computed using Tsiolkovsky's rocket equation \cite{tsiolkovsky1903}:
\begin{equation*}
\begin{split}
\Delta v_{\text{analytical}} &= g_0\, I_{sp}\, \ln\!\left(\frac{m_0}{m_0-m_p}\right) \\
&\quad = 9.81~\mathrm{m\,s^{-2}} \cdot 7{,}377\,\text{s}\cdot \ln\!\left(\frac{1{,}700.430\,\text{kg}}{1{,}700.430\,\text{kg}-478.245\,\text{kg}}\right) \\
&\quad \approx 23{,}899~\mathrm{m\,s^{-1}}.
\end{split}
\end{equation*}

This result is compared with the simulation output, where the velocity increment is obtained by numerically integrating the thrust acceleration:
\begin{equation*}
\Delta v_{\text{sim}} = \int_{0}^{t_{\text{final}}} \frac{\lvert F_T(t) \rvert}{m_{sc}(t)}\,dt
\approx \sum_{i=0}^{n} \frac{\lvert F_T(t_i) \rvert}{m_{sc}(t_i)}\,\Delta t_i
\approx 23{,}899~\mathrm{m\,s^{-1}}.
\end{equation*}

Thus, the absolute error is given by
\[
\left|\Delta v_{\text{analytical}} - \Delta v_{\text{sim}}\right| < 1~\mathrm{m\,s^{-1}},
\]
confirming that the numerical integration reproduces the analytical $\Delta v$ to within $1~\mathrm{m\,s^{-1}}$.

%%%%%%%%%%%%%%

\subsection{Optimisation Reference Case Without Jupiter Gravity Assist}\label{app:nojga_case}

\subsubsection{Trajectory}\label{app:trajectory_without_JGA}

\begin{figure*}[t!]
  \centering
  \includegraphics[width=\textwidth]{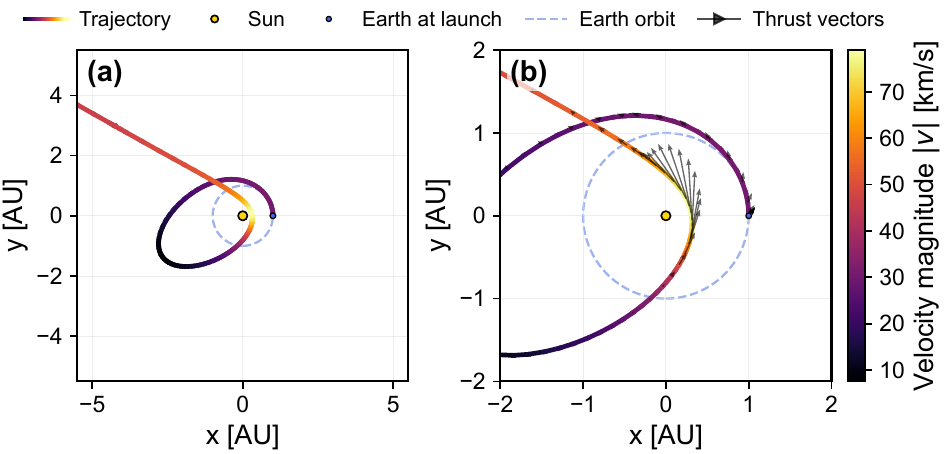}
  \caption{\textbf{Optimised SEP--SOM trajectory without Jupiter gravity assist.}
  Trajectory in the heliocentric in-plane frame; marker colour encodes the heliocentric \emph{velocity magnitude} $|v|$ (colourbar).
  The dashed circle indicates the Earth orbit; the blue marker indicates Earth at launch and the yellow disk denotes the Sun.
  a, Full in-plane trajectory. b, Zoomed view of the near-Sun arc. SEP, solar electric propulsion; SOM, solar Oberth maneuver.}
  \label{fig:traj_combined_no_JGA}
\end{figure*}

Fig.~\ref{fig:traj_combined_no_JGA} shows the trajectory without JGA.

%%%%%%%%%%%%%%

\subsubsection{Learned Strategy}
\label{app:learned_strategy_no_JGA}

\begin{figure}[!t]
    \centering
    \includegraphics[width=\columnwidth]{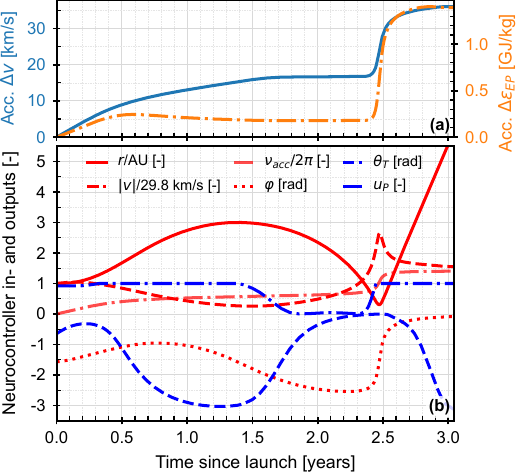}
    \caption{\textbf{Learned steering profile for the SOM trajectory without a Jupiter gravity assist.} Upper panel: cumulative EP $\Delta v$ and EP-induced specific orbital energy gain. Lower panel: controller inputs and outputs.}
    \label{fig:neuro_nojga_io}
\end{figure}

Figure~\ref{fig:neuro_nojga_io} shows the learned steering policy for the
SEP--SOM trajectory without a Jupiter gravity assist. The neurocontroller
maps the state to controls according to the policy encoded by its neural
network genome \(\xi\), yielding thrust angle \(\theta_T\) and throttle
\(u_P\) over mission time \(t\). The upper panel tracks the cumulative
electric-propulsion \(\Delta v\) and the corresponding EP-induced specific
orbital energy gain \(\Delta\varepsilon_{\mathrm{EP}}\). The lower panel
shows selected controller inputs (red) as well as outputs (blue), namely \(\theta_T\) and
\(u_P\).

The controller maintains a power throttle \(u_P=1\) and a thrust angle of \(0^\circ\) during the solar Oberth maneuver. The policy thus concentrates electric-propulsion work in the high-velocity arc around perihelion, consistent with the continuous Oberth effect relation in
Eq.~\eqref{eq:oberth_cont_general}.

By inspecting thrust direction and throttle level in
Fig.~\ref{fig:neuro_nojga_io}, the qualitative phase structure remains
similar to the JGA case. Five phases can be identified:
\begin{enumerate}
 \item \textbf{Aphelion raise (0.0--0.6~yr).} Full throttle, prograde pointing to raise aphelion.
 \item \textbf{Perihelion decrease (0.6--1.8~yr).} Near-full throttle, retrograde pointing to drive perihelion to the design value.
 \item \textbf{Coast (1.8--2.3~yr).} Throttle near zero to set up the solar passage.
 \item \textbf{Solar Oberth (2.3--2.55~yr).} Full-power, prograde thrust through perihelion, consistent with the continuous Oberth energy accumulation in Eq.~\eqref{eq:oberth_cont_general}.
 \item \textbf{Post-SOM outbound arc (2.55--3.0~yr).} A follow-on thrust segment. In the no-JGA case, the controller maintains \(u_P \approx 1\) directly after perihelion, thereby extending high-power thrusting into the post-perihelion arc.
\end{enumerate}

A qualitative comparison to the JGA case in Fig.~\ref{fig:neuro_jga_io}
indicates that the main difference lies in the post-perihelion throttle
behaviour. The no-JGA solution holds \(u_P = 1\) continuously through the
post-perihelion outbound arc, with no throttle reduction, whereas the
JGA solution shows two minor \(u_P\) dips after perihelion. Three
explanations are plausible. First, in the JGA case, the SEP--SOM also
shapes the subsequent Jupiter approach geometry, which affects the
effectiveness of the gravity assist. The energy gain of the JGA depends
on the angle of approach \cite{MesserschmidFasoulas2017_Raumfahrtsysteme}.
Hence a strictly local energy gain right after perihelion from the higher
velocity may be traded for a favourable post-SOM velocity vector and
encounter geometry. In the first dip, only \(u_P\)
deviates slightly from unity, while \(\theta_T\) remains close to
\(0~\mathrm{rad}\). Any significant deviations of \(\theta_T\) would lead
to sub-optimal usage of propellant due to sub-optimal energy gain, as
indicated by Eq.~\eqref{eq:oberth_cont_general}. Second, since the JGA
provides additional \(\Delta v\), propellant can be saved to increase
payload mass, which can manifest as \(u_P < 1\). Third, the variation may
reflect minor sub-optimality from the evolutionary tuning of \(\xi\).

%%%%%%%%%%
\section{Supplementary material: Detailed Derivation of the Generalisation Equation}
\label{app:gen_derivation}

\subsection{Setup and first principle}

We consider heliocentric motion with thrust:
\begin{equation}
  \ddot{\mathbf{r}}(t)
  =
  -\frac{GM_\odot}{r(t)^3}\,\mathbf{r}(t)
  +
  \frac{F_T(t)}{m(t)}\,\hat{\mathbf{t}}(t),
  \label{eq:eom}
\end{equation}
where $m(t)$ is the spacecraft mass, $F_T(t)$ the thrust magnitude, and $\hat{\mathbf{t}}(t)$ the thrust-direction unit vector (parameterised by the thrust angle $\theta_T(t)$).
For two designs started from identical initial conditions to follow the \emph{same} trajectory $\mathbf{r}(t)$, the commanded thrust direction and thrust-acceleration history must match pointwise. A convenient statement of this requirement is
\begin{equation}
  \hat{\mathbf{t}}(t)\ \text{identical},
  \qquad
  \frac{F_T(t)}{m(t)}\ \text{identical}\quad \forall\,t.
  \label{eq:same_acc_history}
\end{equation}

\subsection{Power--thrust--mass model}

Let the EPS deliver power according to a $1/r^\kappa$ law with throttle $u_P(t)\in[0,1]$:
\begin{equation}
  f(t) \equiv \Bigl(\tfrac{1\,\mathrm{AU}}{r(t)}\Bigr)^{\kappa},
  \qquad
  P(t) = \alpha_{\mathrm{EPS}}\,m_{\mathrm{EPS}}\,f(t)\,u_P(t),
  \label{eq:power_model}
\end{equation}
where $\alpha_{\mathrm{EPS}}$ is the EPS specific power at $1$\,AU and $m_{\mathrm{EPS}}=\mu_{\mathrm{EPS}}\,m_0$ is the EPS mass (with $m_0$ the launch mass).
Assuming constant exhaust velocity $c_e$ and thrust efficiency $\eta$ (taken identical for the compared designs),
\begin{equation}
\begin{aligned}
  F_T(t) &= \frac{2\eta}{c_e}\,P(t),\\
  \dot m(t) &= -\,\frac{F_T(t)}{c_e}
  = -\,\frac{2\eta}{c_e^{2}}\,\alpha_{\mathrm{EPS}}\,m_{\mathrm{EPS}}\,f(t)\,u_P(t).
\end{aligned}
  \label{eq:T_and_mdot}
\end{equation}
Define the throttled-irradiance integral
\begin{equation}
  G(t) \equiv \int_{0}^{t} f(t')\,u_P(t')\,\mathrm{d}t'.
  \label{eq:G}
\end{equation}
Integrating \eqref{eq:T_and_mdot} gives
\begin{equation}
  m(t) = m_0 - \frac{2\eta}{c_e^{2}}\,\alpha_{\mathrm{EPS}}\,m_{\mathrm{EPS}}\,G(t),
  \label{eq:m_of_t}
\end{equation}
and the thrust-acceleration magnitude
\begin{equation}
  \frac{F_T(t)}{m(t)} =
  \frac{ \tfrac{2\eta}{c_e}\,\alpha_{\mathrm{EPS}}\,m_{\mathrm{EPS}}\,f(t)\,u_P(t)}
       { m_0 - \tfrac{2\eta}{c_e^{2}}\,\alpha_{\mathrm{EPS}}\,m_{\mathrm{EPS}}\,G(t)}.
  \label{eq:accel_law}
\end{equation}

\subsection{Trajectory-preserving invariant (allowing variable launch mass)}

Consider two designs, ``sim'' (reference) and ``new'', with potentially different launch masses $m_{0,\mathrm{sim}}$ and $m_{0,\mathrm{new}}$, but the same $\eta$, $c_e$, and pointing history $\hat{\mathbf{t}}(t)$.
Substituting $m_{\mathrm{EPS}}=\mu_{\mathrm{EPS}}\,m_0$ into \eqref{eq:accel_law} yields
\begin{equation}
  \frac{F_T(t)}{m(t)} =
  \frac{ \tfrac{2\eta}{c_e}\,\alpha_{\mathrm{EPS}}\,\mu_{\mathrm{EPS}}\,m_0\,f(t)\,u_P(t)}
       { m_0 - \tfrac{2\eta}{c_e^{2}}\,\alpha_{\mathrm{EPS}}\,\mu_{\mathrm{EPS}}\,m_0\,G(t)}
  =
  \frac{ \tfrac{2\eta}{c_e}\,\alpha_{\mathrm{EPS}}\,\mu_{\mathrm{EPS}}\,f(t)\,u_P(t)}
       { 1 - \tfrac{2\eta}{c_e^{2}}\,\alpha_{\mathrm{EPS}}\,\mu_{\mathrm{EPS}}\,G(t)}.
  \label{eq:accel_reduced}
\end{equation}
In this reduced form, the explicit factor $m_0$ cancels; the acceleration history is controlled by $\alpha_{\mathrm{EPS}}\mu_{\mathrm{EPS}}$ together with the throttle history $u_P(t)$ (and the induced history term $G(t)$).

A practical trajectory-preserving condition is the pointwise equality
\begin{equation}
  \alpha_{\mathrm{EPS}}\,\mu_{\mathrm{EPS}}\,u_P(t)
  =
  \alpha_{\mathrm{EPS,sim}}\,\mu_{\mathrm{EPS,sim}}\,u_{P,\mathrm{sim}}(t)
  \quad \forall\,t,
  \label{eq:invariance_mu}
\end{equation}
which ensures identical thrust-acceleration histories through \eqref{eq:accel_reduced} under the stated assumptions.

Two useful corollaries follow:

\subsubsection{(i) Unchanged throttle}
If $u_P(t)=u_{P,\mathrm{sim}}(t)$ is retained, then
\begin{equation}
  \alpha_{\mathrm{EPS}}\,\mu_{\mathrm{EPS}}
  =
  \alpha_{\mathrm{EPS,sim}}\,\mu_{\mathrm{EPS,sim}}.
  \label{eq:alpha_mu_const}
\end{equation}
This expresses a scale-invariant trade between EPS specific power and EPS mass fraction for a fixed throttle schedule.

\subsubsection{(ii) Throttle compensation}
If $\alpha_{\mathrm{EPS}}\mu_{\mathrm{EPS}}$ changes, the same trajectory can still be targeted by adjusting the throttle according to
\begin{equation}
  u_P(t)
  =
  u_{P,\mathrm{sim}}(t)\,
  \frac{\alpha_{\mathrm{EPS,sim}}\,\mu_{\mathrm{EPS,sim}}}
       {\alpha_{\mathrm{EPS}}\,\mu_{\mathrm{EPS}}},
  \qquad
  0\le u_P(t)\le 1\ \ \forall t.
  \label{eq:throttle_comp}
\end{equation}

\subsection{Mass budget and payload expression}

Let $\mu_{\mathrm{p}}\equiv m_{\mathrm{prop}}(0)/m_0$, $\mu_{\mathrm{s}}\equiv m_{\mathrm{s}}/m_0$, and $\mu_{\mathrm{pl}}\equiv m_{\mathrm{pl}}/m_0$.
Under the trajectory-preserving condition \eqref{eq:invariance_mu}, the normalised mass history implied by \eqref{eq:m_of_t} is the same, and in particular the required propellant mass fraction $\mu_{\mathrm{p}}$ is unchanged between the compared cases (for a common terminal time).
Hence the mass budget in terms of fractions is
\begin{equation}
  \mu_{\mathrm{pl}} = (1-\mu_{\mathrm{p}}) - \mu_{\mathrm{s}} - \mu_{\mathrm{EPS}}.
  \label{eq:mass_budget}
\end{equation}
Under the unchanged-throttle invariant \eqref{eq:alpha_mu_const}, the EPS mass fraction satisfies
\begin{equation}
  \mu_{\mathrm{EPS}}
  =
  \frac{\alpha_{\mathrm{EPS,sim}}\,\mu_{\mathrm{EPS,sim}}}{\alpha_{\mathrm{EPS}}}.
  \label{eq:mu_eps_from_invariant}
\end{equation}
Substituting \eqref{eq:mu_eps_from_invariant} into \eqref{eq:mass_budget} gives
\begin{equation}
  \mu_{\mathrm{pl}}
  =
  (1-\mu_{\mathrm{p}}) - \mu_{\mathrm{s}} - \frac{\alpha_{\mathrm{EPS,sim}}\,\mu_{\mathrm{EPS,sim}}}{\alpha_{\mathrm{EPS}}}.
  \label{eq:mu_pl_final}
\end{equation}
Multiplying by the launch mass of the ``new'' case yields the payload mass:
\begin{equation}
  m_{\mathrm{pl,new}} =
  m_{0,\mathrm{new}}
  \left[
    (1-\mu_{\mathrm{p}}) - \mu_{\mathrm{s}} - \frac{\alpha_{\mathrm{EPS,sim}}\,\mu_{\mathrm{EPS,sim}}}{\alpha_{\mathrm{EPS}}}
  \right].
  \label{eq:m_pl_final}
\end{equation}

\subsubsection{Remarks}
\begin{enumerate}
    \item[(i)] The invariance assumes identical $c_e$ and $\eta$, and identical pointing histories $\hat{\mathbf{t}}(t)$.
    \item[(ii)] If throttle compensation \eqref{eq:throttle_comp} saturates (i.e., $u_P(t)>1$ at some $t$), the same trajectory cannot be maintained with that $(\alpha_{\mathrm{EPS}},\mu_{\mathrm{EPS}})$ pair under the present model.
    \item[(iii)] The exponent $\kappa$ in $f(t)$ is arbitrary for this derivation; it cancels in the invariance because $f(t)$ is common when $\mathbf{r}(t)$ is identical.
\end{enumerate}


%%    \citept{<label>} ==> Jones et al. [21]
%%    \cite{<label>} ==> [21]
%%

\phantomsection  % Anchors the link to the top of this page
\addcontentsline{toc}{chapter}{References} % Adds to TOC and Bookmarks
\bibliographystyle{elsarticle-num-names} 
\bibliography{references/overall_references}





% \end{thebibliography}
\end{document}